\documentclass[10pt]{article}

\usepackage{ragged2e}       
\usepackage{soul}           
\usepackage{xcolor}
\usepackage{graphicx}
\usepackage{hyperref}
\usepackage{amsmath}
\usepackage{subcaption}
\usepackage{amssymb}
\usepackage{booktabs}
\usepackage{fullpage}
\usepackage[section]{placeins}
\usepackage{authblk}

\definecolor{my_purple}{RGB}{255,0,255}
\graphicspath{{LoopFigures/}}

\definecolor{REDCOLOR2}{RGB}{0,0,0}

\title{\textbf{\Large Pressure – area loop based phenotypic classification and mechanics of esophagogastric junction physiology}}

\author[1]{\normalsize Guy Elisha}
\author[2]{\normalsize Sourav Halder}
\author[1]{\normalsize Shashank Acharya}
\author[3]{\normalsize Dustin A. Carlson}
\author[3]{\normalsize Wenjun Kou}
\author[3]{\normalsize Peter J. Kahrilas}
\author[3]{\normalsize John E. Pandolfino}
\author[1,2]{\normalsize Neelesh A. Patankar\thanks{Corresponding author: N.~A.~Patankar (\texttt{n-patankar@northwestern.edu})}}

\affil[1]{Department of Mechanical Engineering, McCormick School of Engineering, \newline
Northwestern University Technological Institute, 2145 Sheridan Road, Evanston, IL 60201 \vspace{1ex}}
\affil[2]{Theoretical and Applied Mechanics Program, McCormick School of Engineering,\newline
Northwestern University Technological Institute, 2145 Sheridan Road, Evanston, IL 60201 \vspace{1ex}}
\affil[3]{Division of Gastroenterology and Hepatology, Feinberg School of Medicine, \newline
Northwestern University, 676 North St. Clair Street, Arkes Suite 2330, Chicago, IL 60611 \vspace{1ex}}

\date{}

\begin{document}

\maketitle 

\begin{abstract}
The esophagogastric junction (EGJ) is located at the distal end of the esophagus and acts as a valve allowing swallowed materials to enter the stomach and preventing acid reflux. Irregular weakening or stiffening of the EGJ muscles result in changes to its opening and closing patterns which can progress into esophageal disorders. Therefore, understanding the physics behind the opening and closing cycle of the EGJ provides a mechanistic insight into its function and can help identify the underlying conditions that cause its degradation. Using clinical FLIP data, we plotted the pressure-area hysteresis at the EGJ location and distinguished two major loop types, a pressure dominant loop (PDL) and a tone dominant loop (TDL). In this study, we aimed to identify the key characteristics that define each loop type and find what causes the inversion from one loop to another. To do so, the clinical observations were reproduced using 1D simulations of flow inside a FLIP device located in the esophagus, and the work done by the EGJ wall over time was calculated. This work was decomposed into active and passive components, which revealed the competing mechanisms that dictate the loop type. These mechanisms are esophagus stiffness, fluid viscosity, and the EGJ relaxation pattern. In PDL, the leading source of energy in the cycle is coming from the fluid pressure increase from the peristaltic contraction wave, and in TDL the leading source of energy in the cycle is coming from the contraction and relaxation of the EGJ tone.
\end{abstract}


\section{Introduction}

The esophagus is a tubular organ that connects between the mouth and the stomach and is responsible for transferring swallowed food towards the stomach through a peristaltic contraction wave (muscle contraction) \cite{Mittal2016}. The esophagogastric junction (EGJ) is a neurally controlled valve located at the distal end of the esophagus. It is an important component of the digestive track, as it monitors the entering of material coming from the mouth to the stomach while preventing gastric acid from the stomach to enter the esophagus \cite{Mittal1997}. 

The degradation of the muscles at the EGJ can result in a more compliant EGJ, causing acid reflux which can damage the esophagus \cite{Pandolfino2003,Kwiatek2010}. This damage, caused by some degree of uncontrolled opening of the EGJ, can lead to several digestive disorders which can progress into esophageal diseases such as Gastroesophageal reflux disease (GERD) \cite{Pandolfino2003,Kahrilas2008}. It has been reported that GERD is the most common gastrointestinal diagnosis during doctors' office visits, and that 14-20\% of the adult population in the United States suffers from some degree of GERD \cite{Kahrilas2008,Shaheen2006}.
On the other end of the spectrum, diseases such as Achalasia are characterized by a lack of peristalsis and an absence of EGJ relaxation which restricts EGJ opening and emptying of the esophagus, preventing a successful swallow from occurring \cite{Boeckxstaens2014,Eckardt2009}. 

Therefore, understanding the physics behind the opening and closing of the EGJ provides a mechanistic insight into its functions and can help identify the underlying conditions that cause its degradation. In this work, we take a mechanistic approach to explore the opening and closing paths of the EGJ through looking at how energy is distributed at the EGJ during a contractile cycle. Doing so helps to identify the mechanism and physical parameters that characterize the EGJ functions.

\subsection{Background} \label{Background}

One tool that is used today to determine the shape of the esophagus is the functional luminal imaging probe (FLIP), a balloon catheter which measures the cross sectional-area of the esophagus at 16 locations and one distal pressure measurement as a function of time \cite{Carlson2015,Carlson2016}. The FLIP is inserted into the esophagus, placed within the esophageal lumen. The FLIP bag is incrementally inflated with fluid and the esophagus wall responds to it by contracting and relaxing its muscles in order to push fluid towards the distal end of the esophagus. As a result of the moving contraction, the pressure at the distal end increases. 

\begin{figure*}
    \centering{{\includegraphics[trim=0 0 0 0 ,clip,width=0.9\textwidth]{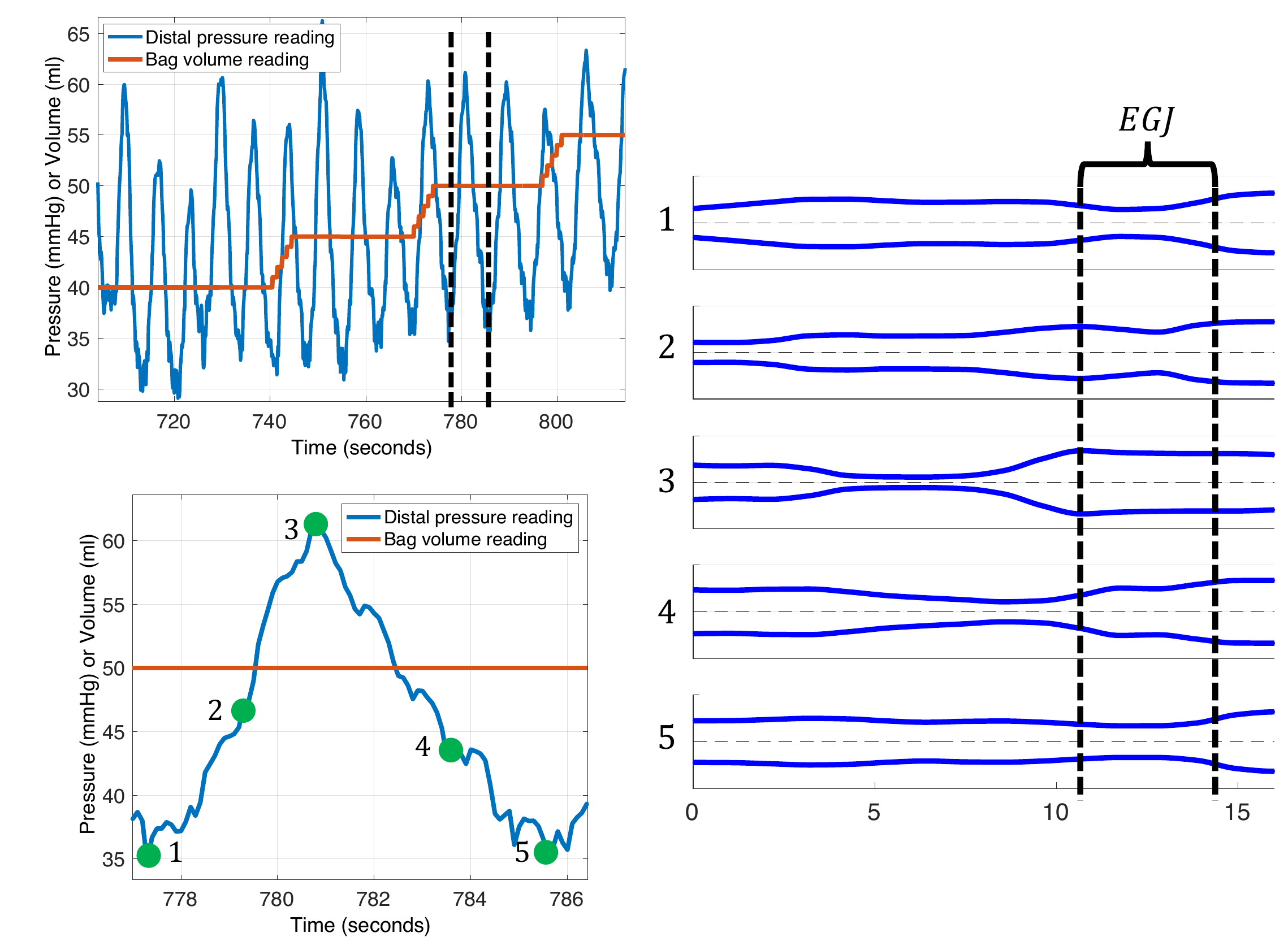}}}
    \caption{Visual representation of clinical data collected from the FLIP. The graph on the top left shows the pressure reading and bag volume over 120 seconds. The graph on the bottom left is a close-up look of the data between the two dotted lines in the plot above it, highlighting a single contractile cycle. The right figure displays the tube shape at five different time instances, ordered chronologically, corresponding to the five points on the pressure graph. }
    \label{fig:FLIP_figure}
\end{figure*}

Figure \ref{fig:FLIP_figure} presents FLIP data recorded from a single subject. The graph on the top left presents the pressure reading and bag volume as a function of time. The pressure changes as a result of esophageal contractility in response to distension. The volume increase is dictated by the physician conducting the procedure. The graph on the bottom left shows a closer look at one particular section from the plot above it (between 777.3 and 785.6 seconds). It captures the pressure variation during a single contractile cycle \cite{Carlson2015}. The five points highlighted on this graph represent five time instances in the contractile cycle. The esophagus shape at each of these instances is displayed on the right of figure \ref{fig:FLIP_figure}, and the location of the EGJ is marked. The EGJ is identified as the location with the minimum cross-sectional area at the beginning of the contractile cycle. As this figure shows, the increase in pressure corresponds to the increase of cross-sectional area at the EGJ, and pressure decrease at the end of the contractile cycle corresponds to the decrease of cross-sectional area at the EGJ. The maximum opening of the EGJ occurs at the same time instance as the maximum pressure.

\subsection{Pressure-Area Loops at EGJ} \label{PressureAreaLoops_Intro}

\begin{figure*}[!htb]
    \centering
    \begin{subfigure}[b]{0.480\textwidth}
        \centering
        {\includegraphics[trim=30 180 60 200,clip,width=\textwidth]{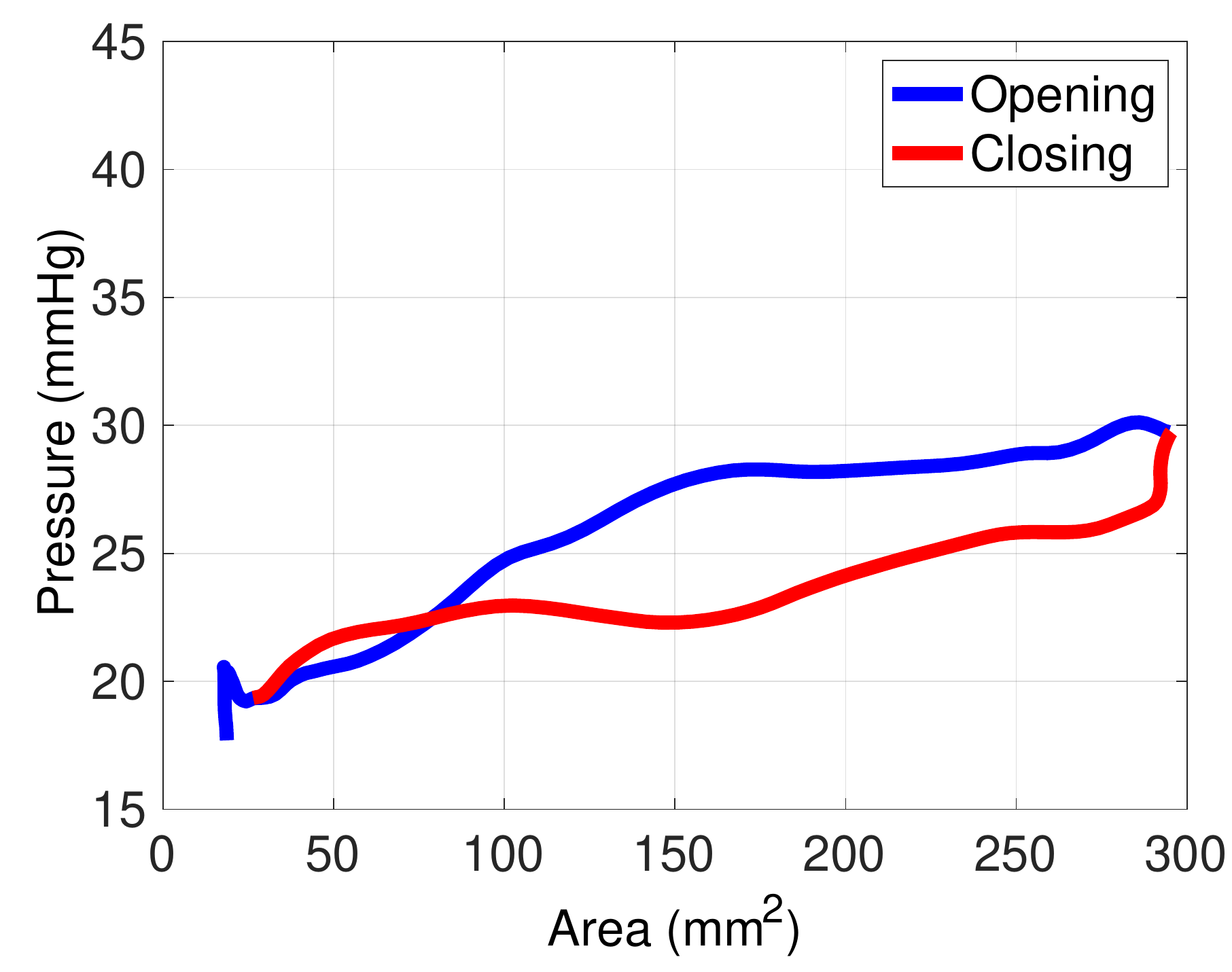}}
        \caption{Pressure dominant loop}
        \label{fig:PDL_clinical}
    \end{subfigure}
    \hfill
    \begin{subfigure}[b]{0.480\textwidth}  
        \centering 
        {\includegraphics[trim=30 180 60 200,clip,width=\textwidth]{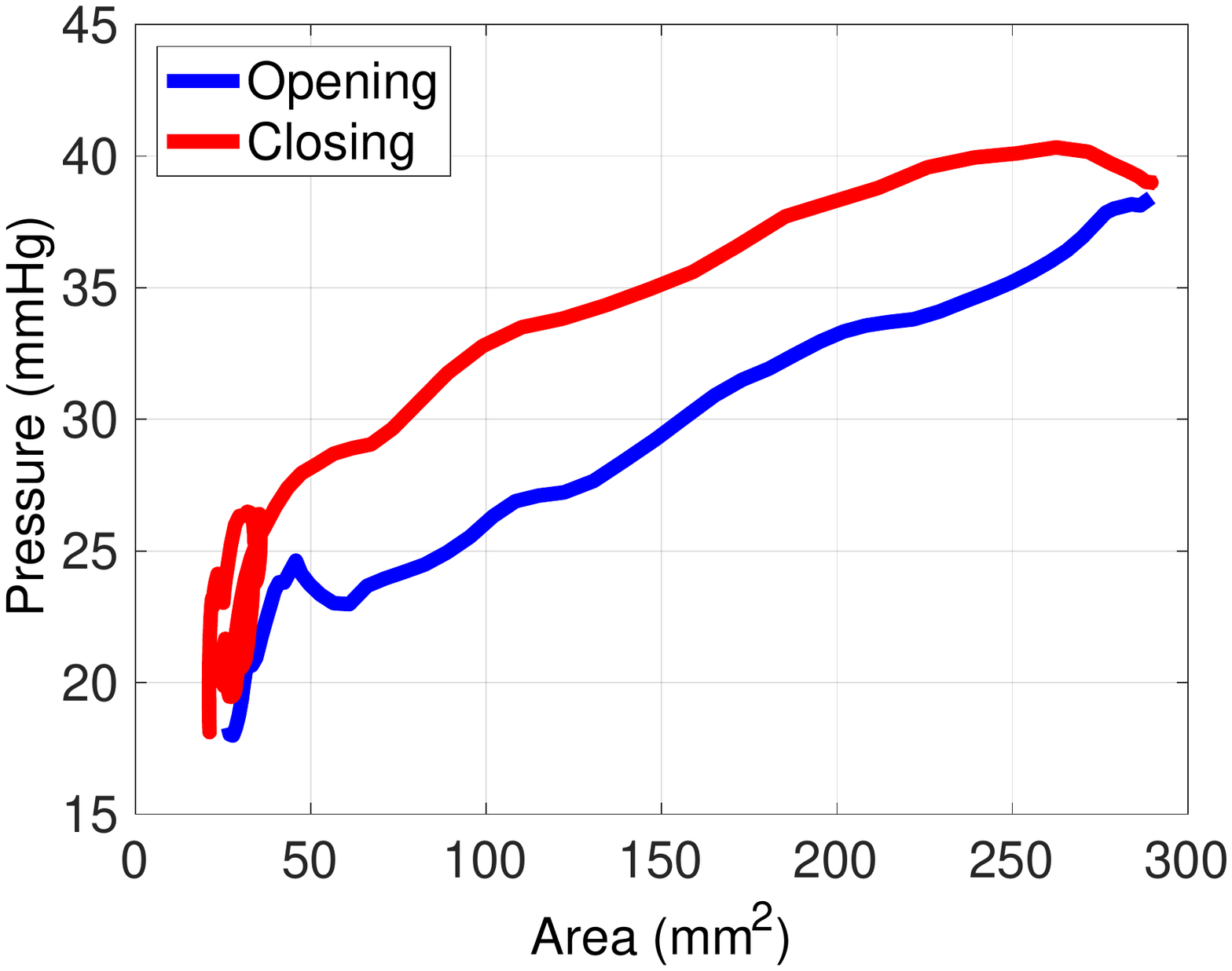}}
        \caption{Tone dominant loop}
        \label{fig:TDL_clinical}
    \end{subfigure}
    \caption{Two clinically observed pressure-area loops at the esophagogastric junction.} 
    \label{fig:clinical_Loops}
\end{figure*}

Figure \ref{fig:PDL_clinical} presents a plot of the pressure at the EGJ as a function of EGJ cross-sectional area during a single contractile cycle. The pressure at the EGJ is obtained by using a constitutive equation for pressure (discussed in section \ref{Governing_Equations}), as proposed by \cite{Halder_2021}. As figure \ref{fig:PDL_clinical} shows, the cross-sectional area at the EGJ increases with pressure increase, and decreases with pressure decrease. This trend is opposite to the one observed in other sphincters such as the upper esophageal sphincter (UES) and the anal canal (AC) \cite{Omari2015,Zifan2019}. The reason for this is that the UES and AC open due to a neurally controlled activation \cite{Omari2015,Zifan2019}, whereas the EGJ opening is controlled by two functions.  The first is a neurally mediated EGJ relaxation (such as for other sphincters), and the second is a pressure rise created by the peristaltic contraction wave which leads to the opening. Therefore, understanding the physics behind the opening and closing pattern of the EGJ entails an additional complication.

The pressure-area plot creates a loop with distinct opening and closing curves (marked in blue and red, respectively), which indicates that there is some energy that is being gained or lost by the system. We examined 192 contractile cycles of 24 randomly selected, asymptomatic volunteers, and plotted their pressure-area loops, identifying the opening and closing curves. The FLIP data used in this study was collected at the Esophageal Center of Northwestern between November 2012 and October 2018, using a 16-cm FLIP (EndoFLIP\textsuperscript{\tiny\textregistered} EF-322N; Medtronic, Inc, Shoreview, MN) \cite{AcharyaEsoWork2020,Carlson2021}. Additional details on the data collection process and subject cohort selection is available in \cite{AcharyaEsoWork2020, Lin2013,Carlson2016,Carlson2021}. We characterize a contractile cycle by at least 3 cm decrease in the luminal diameter and contractions of at least 6cm traveling forward along the length of the esophagus \cite{AcharyaEsoWork2020,Carlson2021}. By making the distinction between the opening and closing curves, we identified two, equally appearing, physically meaningful loop types. The first loop type, displayed in figure \ref{fig:PDL_clinical} is a regular hysteresis, where the opening curve is above the closing curve. In contractile cycles with this loop type, the dominant source of energy in the opening and closing cycle of the EGJ is the contraction wave. This loop is called pressure dominant loop (PDL). The second loop type, displayed in figure \ref{fig:TDL_clinical} is flipped, where the closing curve is above the opening curve. In contractile cycles with this loop type, the dominant source of energy in the opening and closing cycle of the EGJ is the active relaxation and contraction of the EGJ tone. This loop is called tone dominant loop (TDL). The reason behind these names is discussed later in this writing.

\subsection{Research Objectives} \label{Objectives}

The goal of this work is to obtain a better understanding of the opening and closing mechanisms of the EGJ. The research objectives are: (i) identify the key characteristics that define each loop type, (ii) find what causes the inversion from one loop to another, and (iii) examine how energy is spent in the EGJ during a single contractile cycle.

\section{Problem Formulation \& Numerical Solution} \label{MathDetails}

\subsection{Governing Equations in 1D} \label{Governing_Equations}

To determine the key parameters that dictate the loop type, we developed a 1D model of a flow inside a FLIP device placed in the esophagus. The model is similar to the one described by \cite{Acharya_2021,Elisha2021}. The mass and momentum conservation equations are 

\begin{equation} \label{eq:continuity}
    \frac{\partial A}{\partial t}+\frac{\partial\left(Au\right)}{\partial x} = 0,
\end{equation}
and
\begin{equation} \label{eq:momentum}
    \frac{\partial u}{\partial t} + u\frac{\partial u}{\partial x} = 
    -\frac{1}{\rho}\frac{\partial P}{\partial x}-\frac{8\pi\mu u}{\rho A},
\end{equation}

\noindent  respectively. In the equations above, $A(x,t)$, $u(x,t)$, and $P(x,t)$ are the tube cross-sectional area, fluid velocity (averaged at each cross-sectional area), and pressure inside the tube, respectively.  The constants $\rho$ and $\mu$ are the fluid density and viscosity, respectively. 
These 1D forms of the continuity and the momentum conservation equations were derived in \cite{Ottesen2003} and have been widely used to describe valveless pumping \cite{Acharya_2021,Manopoulos2006, Bringley2008,Elisha2021}. 

The last equation in this system expresses pressure in terms of the tube's cross-sectional area. This constitutive relation, derived in \cite{Whittaker2010} and validated experimentally in \cite{Kwiatek2011} is known as the 'tube law', and takes the form

\begin{equation}
{\Delta{P}}={K_{\scriptscriptstyle e}}\left(\frac{A(x,t)}{A_{\scriptscriptstyle o}\theta(x,t)}-1\right).
\label{eqn:tube_law_delta}
\end{equation}

\noindent Here, $\Delta{P}$ is the difference between the pressure inside and outside the tube ($\Delta{P}=P_i-P_o$), $K_e$ is tube stiffness, and $A_o$ is the undeformed reference area representing the cross-sectional area of the tube when ${\Delta{P}}=0$ \cite{Acharya_2021,Elisha2021}. Lastly, $\theta(x,t)$ is an activation term which changes the reference cross-sectional area of the tube wall. This term is implemented in order to mimic the function of the esophagus' muscle fibers as they contract and relax throughout the contractile cycle \cite{Acharya_2021,Ottesen2003,Manopoulos2006,Bringley2008,Mittal2016,AbrahaoJr2010,Elisha2021}. The activation term $\theta(x,t)$ is expressed as a piecewise function elaborated upon in section \ref{Peristaltic_Wave}.

Since the outside pressure is constant, equation (\ref{eqn:tube_law_delta}) can be written as

\begin{equation}
{P(x,t)}={K_{\scriptscriptstyle e}}\left(\frac{A(x,t)}{A_{\scriptscriptstyle o}\theta(x,t)}-1\right)+P_o
\label{eqn:tube_law}
\end{equation}

\noindent to solve for the pressure inside the tube. Note that in our simulations, the pressure outside of the tube is assumed to be the reference pressure, so that $P_o=0$.

\subsubsection{Non-dimensionalizing Dynamic Equations} \label{nondimParametersGov}

To obtain a better understanding of the physical held of the system and reduce the number of independent variables, we non-dimetionalize equations (\ref{eq:continuity}), (\ref{eq:momentum}) and (\ref{eqn:tube_law}) using

\begin{equation}
A=\alpha A_{o}, \qquad t=\tau\frac{L}{c}, \qquad u=Uc, \qquad P=pK_e, \qquad \text{and} \qquad x=\chi L,
\end{equation}

\noindent where $\alpha$, $\tau$, $U$, $p$, and $\chi$ are non-dimensional variables of area, time, velocity, pressure, and position, respectively \cite{Acharya_2021,Elisha2021}. The terms $c$ and $L$ are dimensional constants for the speed of the peristaltic wave and the length of the tube, respectively. 

Therefore, the mass conservation, momentum conservation, and tube law equations can be written as
\begin{equation} \label{eq:continuity_nondim}
    \frac{\partial\alpha}{\partial\tau}+\frac{\partial\left(\alpha U\right)}{\partial\chi} = \epsilon\left(\alpha_{xx}\right), 
\end{equation}

\begin{equation} \label{eq:momentum_nondim}
    \frac{\partial U}{\partial\tau} + U\frac{\partial U}{\partial\chi} + 
    \psi\frac{\partial p}{\partial\chi}
    + \beta\frac{U}{\alpha} = 0 , \qquad \text{and}
\end{equation}

\begin{equation} \label{eq:tube_law_nondim}
    \textcolor{REDCOLOR2}{p=\left( \frac{\alpha}{\theta}-1 \right)- f(\alpha,U)},
\end{equation}

\noindent respectively. The terms $\psi$ and $\beta$ are non-dimensional stiffness and viscosity parameters, respectively, defined as $\psi=K_e/(\rho c^2)$ and $\beta = 8\pi\mu L/(\rho A_o c)$. Notice that $\psi$ is inverse of Cauchy number and $\beta$ is inverse of the Reynolds number. The function $f(\alpha,U)$ is a damping term added to the right hand side of the pressure equation to regularize the system and therefore help stabilize the numerical solution. It is defined as $f(\alpha,U)=\eta\frac{\partial\left(\alpha U \right)}{\partial\chi}$ where $\eta=(YcA_o)/(K_eL)$, and Y is the damping coefficient. An additional discussion is available in \cite{Acharya_2021} and \cite{Wang2014}. Lastly, $\epsilon\left(\alpha_{xx}\right)$ is a smoothing term added to the right-hand side of the continuity equation, as in \cite{Acharya_2021}, in order to obtain faster convergence and reduce computational time.

Finally, we can plug equation (\ref{eq:tube_law_nondim}) into equation (\ref{eq:momentum_nondim}) to obtain

\begin{equation} \label{eq:momentum_nondim_final}
\frac{\partial U}{\partial\tau} + U\frac{\partial U}{\partial\chi} + \beta\frac{U}{\alpha} + 
        \psi\frac{\partial}{\partial\chi}\left(\frac{\alpha}{\theta}\right) = 
        \zeta\frac{\partial^2}{\partial\chi^2}\left(\alpha U\right),
\end{equation}
such that we have a system of two equations, equations (\ref{eq:continuity_nondim}), and (\ref{eq:momentum_nondim_final}). Here, $\zeta$ is equal to the product of $\eta$ and $\psi$.

\subsection{Peristaltic Wave Input and Active Relaxation} \label{Peristaltic_Wave}

The activation function $\theta(\chi,\tau)$ multiplies the constant reference area term $A_o$ in the tube law (equation (\ref{eq:tube_law_nondim})) in order to change the reference area to resemble the activation of the esophagus muscle. We apply this function to the model in the intention of mimicking both the muscle contraction of the traveling wave which pushes fluid forward, as well as the contraction and relaxation of the EGJ muscles. Using an activation function to vary the reference area is often used in cases where the external activation pressure at a specific location varies sinusoidally with time \cite{Acharya_2021,Ottesen2003,Manopoulos2006, Bringley2008}.

In \cite{Acharya_2021,Elisha2021}, the activation function only considered the peristaltic contraction of the traveling wave and wall relaxation, but did not include the EGJ muscle contraction and relaxation. In these studies, the muscle activation was modeled as a sinusoidal wave traveling with time along the length of the elastic tube. The wave had constant amplitude, wave speed, and wavelength. This form of the activation function was supported by clinical data reported in \cite{Goyal2008,Crist1984}. The activation function $\theta$ introduced in our model is similar to the ones in \cite{Acharya_2021} and \cite{Elisha2021}, and it is defined as a superposition of two piecewise functions, such that 
\begin{equation} \label{eq:thetaSuperposition}
    \theta(\chi,\tau) = \theta_p(\chi,\tau) + \theta_{\text{EGJ}}(\chi,\tau).
\end{equation}
Function $\theta_{\text{EGJ}}(\chi,\tau)$ is responsible for the active relaxation of the EGJ and function $\theta_p(\chi,\tau)$ is responsible for the traveling peristalsis as well as the contraction of the EGJ tone at the end of the contractile cycle.

The traveling wave activation function $\theta_p(\chi,\tau)$ takes the form

\begin{equation}\label{eq:theta_p}
\theta_p(\chi,\tau)=
  \begin{cases}
          1 - \frac{1-\theta_{c}}{2}  \left[1 \ +  
\sin\left(\frac{2\pi}{\text{w}}\left(\chi-\tau\right) +
        \frac{3\pi}{2}\right)\right], & \tau-\text{w}\leq\chi\leq\tau \text{, }\tau\leq\chi_2 \\
                  1 - \frac{1-\theta_{c}}{2}  \left[1 \ +  
\sin\left(\frac{2\pi}{\text{w}}\left(\chi-\chi_2\right) +
        \frac{3\pi}{2}\right)\right], & \chi_1\leq\chi\leq\chi_2\text{, } \tau>\chi_2 \\
  1, & \text{otherwise}.
\end{cases}
\end{equation}

\begin{figure*}[!htb]
    \centering
    \begin{subfigure}[b]{0.480\textwidth}
        \centering
        {\includegraphics[trim=30 270 40 270,clip,width=\textwidth]{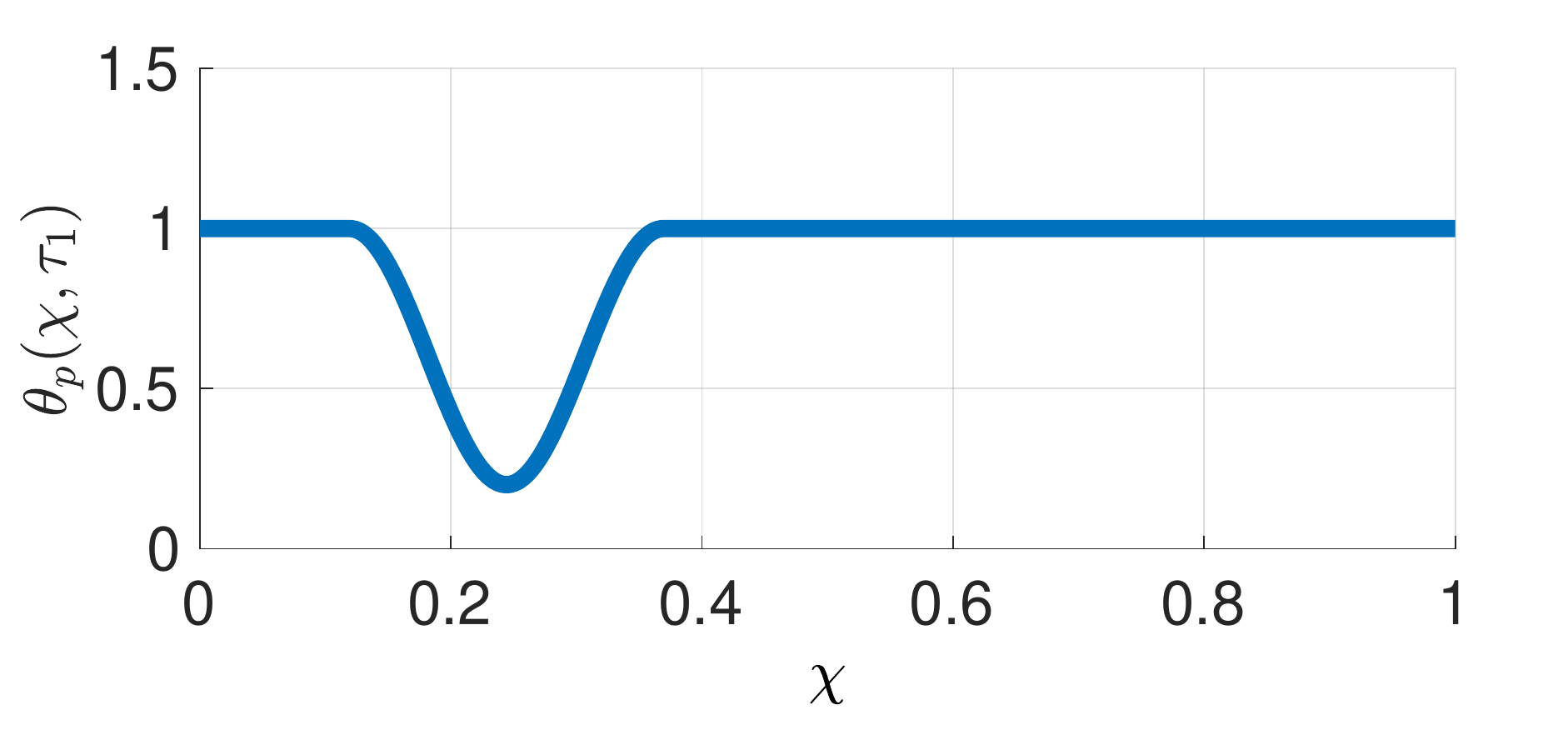}}
        \caption{$\tau=\tau_1<\chi_2$}
        \label{fig:theta_p_t1}
    \end{subfigure}
    \hfill
    \begin{subfigure}[b]{0.480\textwidth}  
        \centering 
        {\includegraphics[trim=30 270 40 270,clip,width=\textwidth]{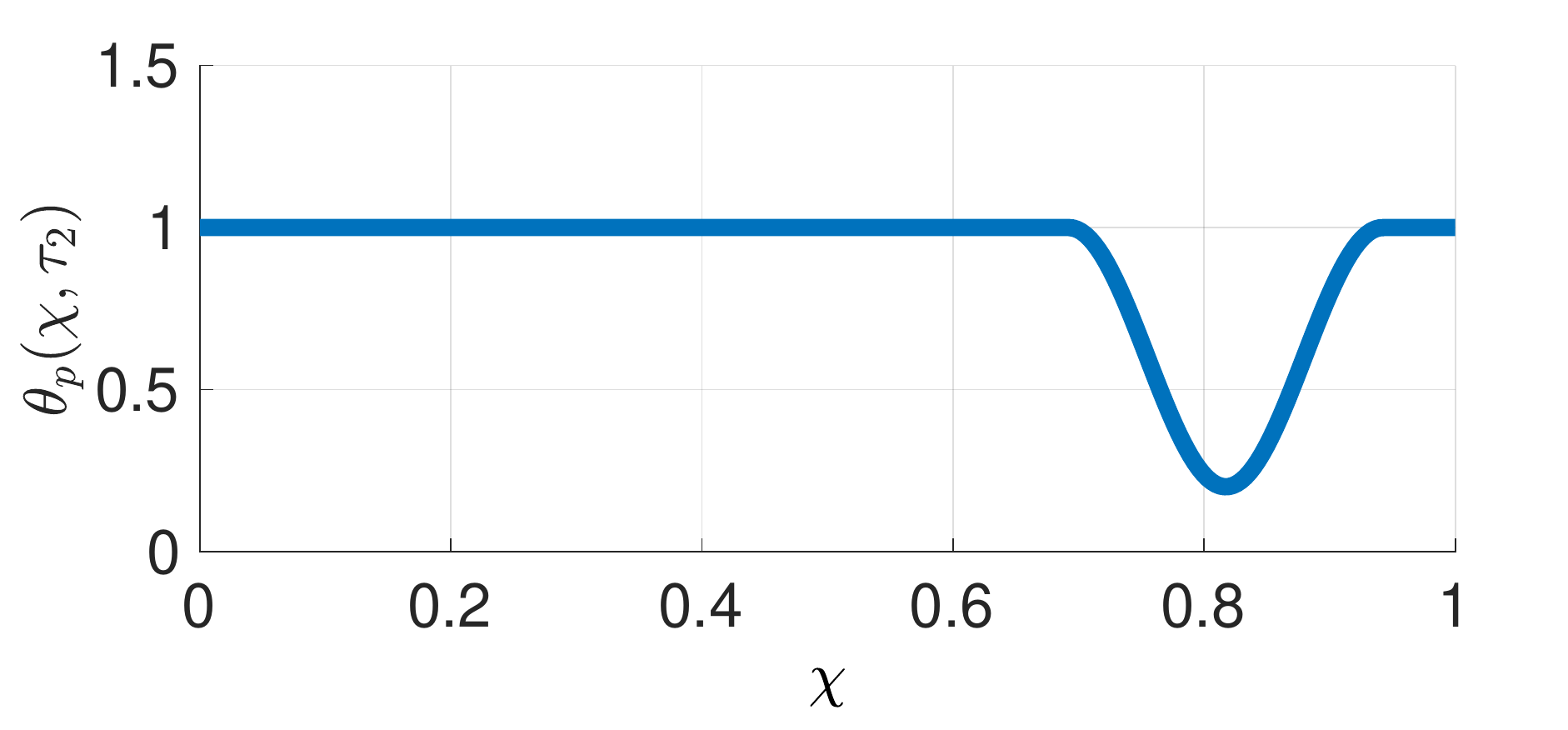}}
        \caption{$\tau=\tau_2>\chi_2$}
        \label{fig:theta_p_t2}
    \end{subfigure}
    \caption{Two plots of function $\theta_p(\chi,\tau)$ defined in equation \ref{eq:theta_p}, at (a) $\tau=\tau_1<\chi_2$ and (b) $\tau=\tau_2>\chi_2$.} 
    \label{fig:theta_p}
\end{figure*}

\noindent In the equation above, $\theta_c$ is the peristaltic contraction strength, $\text{w}$ is the non-dimensional width of the peristaltic wave, and $\chi_1$ and $\chi_2$ are the boundaries of the EGJ segment, where $\chi_2>\chi_1$ and the width of the EGJ is $\text{w}_{\text{EGJ}}=\chi_2-\chi_1$. In this work, $\text{w}_{\text{EGJ}}=\text{w}$. Function $\theta_p$ is plotted at two time instance, $\tau_1$ and $\tau_2$ in figures \ref{fig:theta_p_t1} and \ref{fig:theta_p_t2}, respectively, where $\tau_1<\chi_2$ and $\tau_2>\chi_2$. As the figures show, while $\tau\leq\chi_2$, the wave defined by  $\theta_p(\chi,\tau)$ travels with time along the length of the domain. Once $\tau>\chi_2$, the wave stops traveling in time yet conserves its form (figure \ref{fig:theta_p_t2}).

Function $\theta_{\text{EGJ}}(\chi,\tau)$ is a piecewise function which mimics the initial contraction and active relaxation of the EGJ tone. Different than function $\theta_p$, function $\theta_{\text{EGJ}}$ does not travel with time, but its amplitude decreases with time. The function is defined such that

\begin{equation}\label{eq:theta_EGJ}
\theta_{\text{EGJ}}(\chi,\tau)=
  \begin{cases}
           \frac{\theta_{r}\theta_R(\tau)-1}{2}  \left[1 \ +  
\sin\left(\frac{2\pi}{\text{w}_{\text{EGJ}}}\left(\chi-\chi_2\right) +
        \frac{3\pi}{2}\right)\right], &  \chi_1\leq\chi\leq\chi_2 \\
  0, & \text{otherwise}.
\end{cases}
\end{equation}

\noindent where $\theta_r$ is the EGJ contraction strength and

\begin{equation}\label{eq:theta_R}
\theta_R(\tau)=
  \begin{cases}
          1 , & \tau\leq\tau_i \\
                  1 + \frac{1+\frac{1}{\theta_{r}}}{\tau_m-\tau_i}(\tau-\tau_i),& \tau_i<\tau\leq\tau_m  \\
  \frac{1}{\theta_{r}}, & \tau_m<\tau.
\end{cases}
\end{equation}

\noindent In the equation above, $\tau_i$ marks the time instance in which the EGJ wall starts relaxing, and $\tau_m$ marks the time instance in which the EGJ wall is fully relaxed. We define the relaxation time as $\gamma=\tau_m-\tau_i$. 

As equations (\ref{eq:theta_EGJ}) and (\ref{eq:theta_R}) indicate, at the beginning of the the contractile cycle ($0\leq\tau<\tau_i$), the EGJ is contracted, and therefore the shape of $\theta_{\text{EGJ}}(\chi,\tau=0)$ is similar to $\theta_p$ when $\tau>\chi_2$ (figure \ref{fig:theta_p_t2}). Once $\tau=\tau_i$, the EGJ wall starts relaxing, until $\tau=\tau_m$. At  $\tau>\tau_m$, the EGJ is fully relaxed. Note that the active contraction of the EGJ at the end of the contractile cycle is obtained by the peristaltic contraction wave function $\theta_p$ traveling down the length of the tube and stopping at the EGJ location (stops at $\chi_1\leq\chi\leq\chi_2\text{, } \tau>\chi_2$). For this to occur, we set $\theta_r$=$\theta_c$.

\begin{figure*}[!htb]
    \centering
    \begin{subfigure}[b]{0.480\textwidth}
        \centering
        {\includegraphics[trim=40 80 40 50,clip,width=\textwidth]{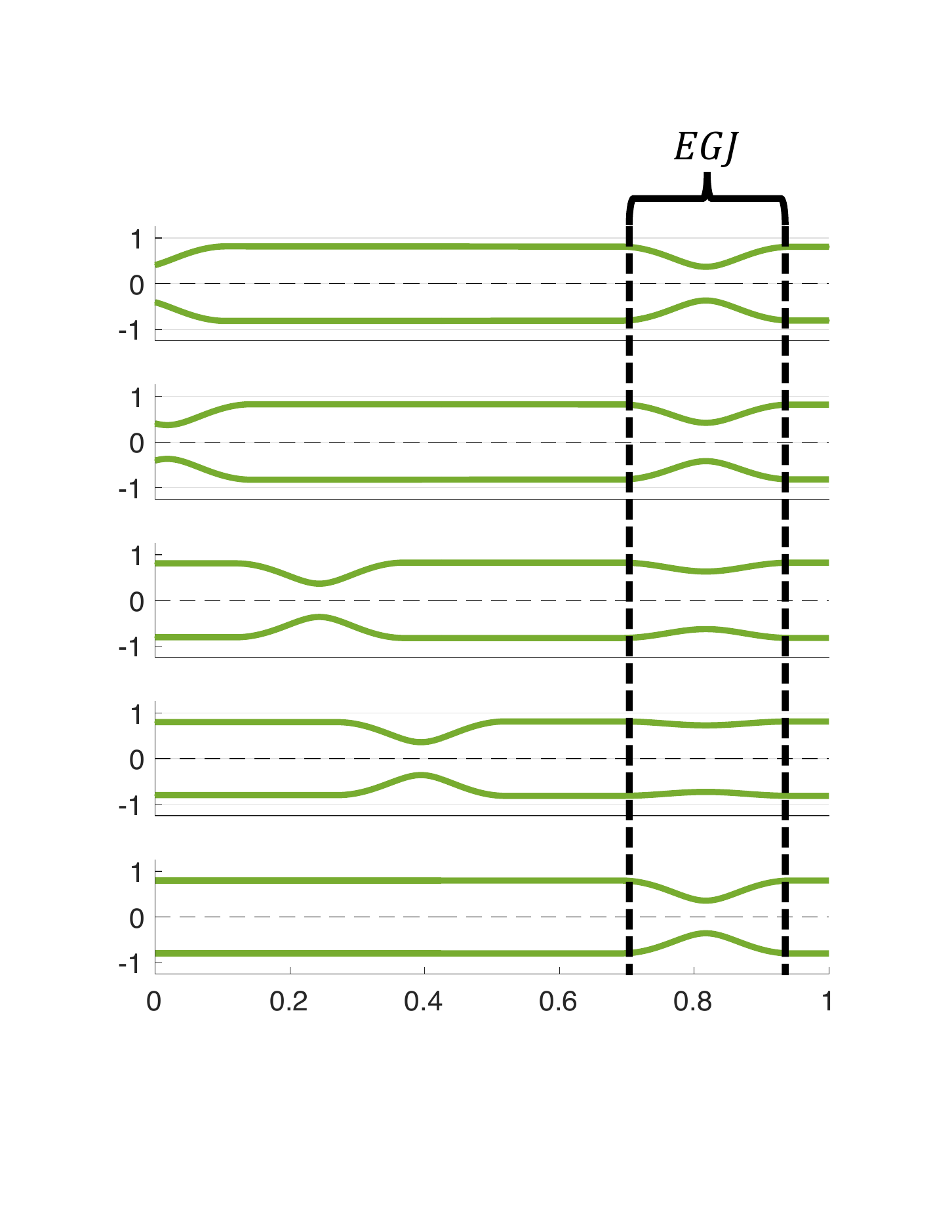}}
        \caption{Slow EGJ wall relaxation}
        \label{fig:Slow_Opening}
    \end{subfigure}
    \hfill
    \begin{subfigure}[b]{0.480\textwidth}  
        \centering 
        {\includegraphics[trim=40 80 40 50,clip,width=\textwidth]{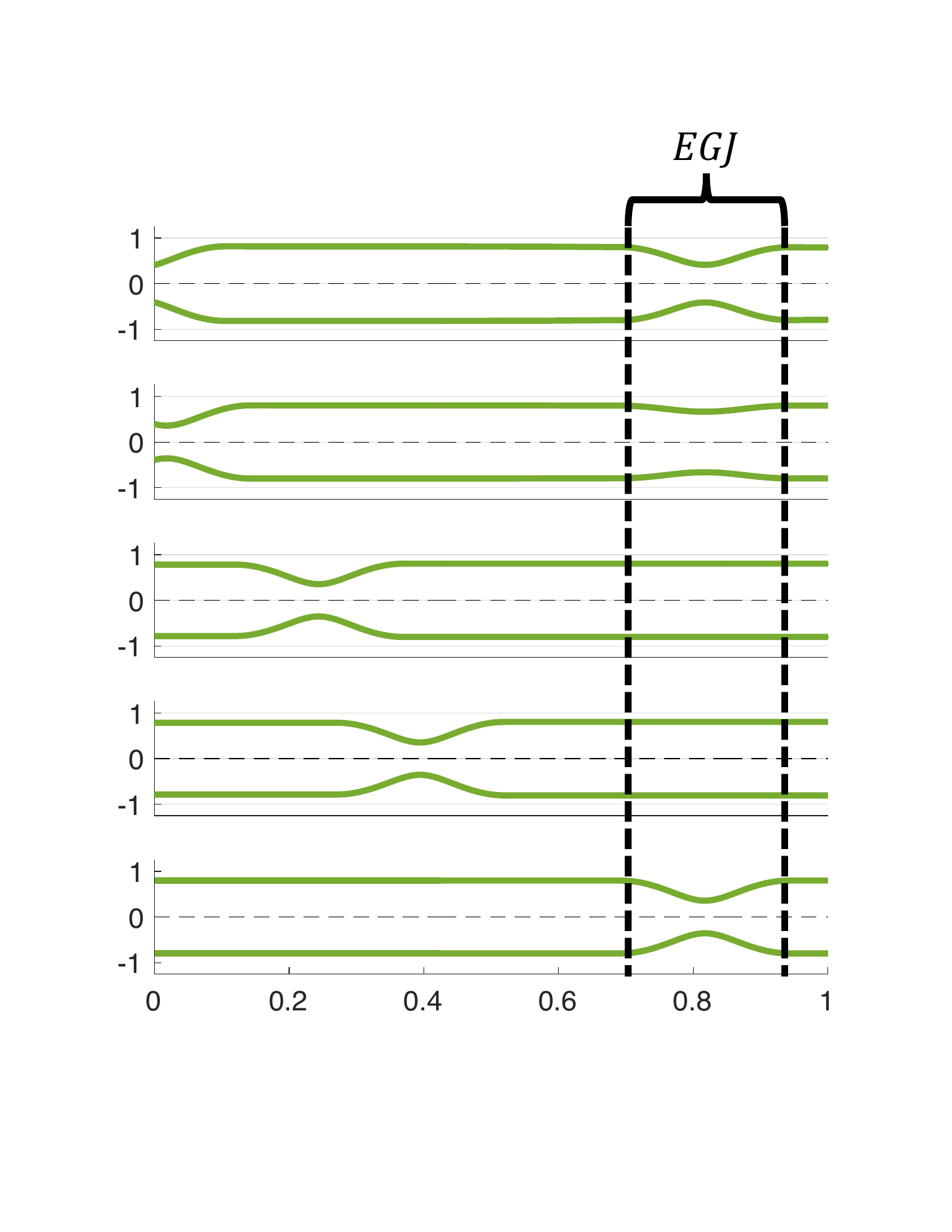}}
        \caption{Fast EGJ wall relaxation}
        \label{fig:Fast_Opening}
    \end{subfigure}
    \caption{Tube deformation of two simulations. (a) EGJ contraction relaxes slow in relation to the peristaltic contraction wave entering the domain. (b) EGJ contraction relaxes fast in relation to the peristaltic contraction wave entering the domain. Both tubes start relaxing at the same time and reach the same maximum relaxation.} 
    \label{fig:Opening_Speed}
\end{figure*}

The value of $\gamma$ dictates the EGJ relaxation speed, which introduces a new physical component into the parameters that define the problem. Small $\gamma$ implies fast relaxation of the EGJ wall and large $\gamma$ implies slow relaxation of the EGJ wall. Figure \ref{fig:Opening_Speed} presents tube deformations of two simulations at five different time instances. Figure \ref{fig:Slow_Opening} displays a case with slow relaxation of the EGJ wall while figure \ref{fig:Fast_Opening} displays a case with fast relaxation of the EGJ wall, such that $\gamma_{a}>\gamma_{b}$. As the two figures show, at the first time instance, the EGJ walls of both tubes are equally contracted. However, at the second time instance, this is no longer the case. While the contraction strength of the EGJ of tube (a) has barely changed from the previous time instance, the EGJ at case (b) is almost fully relaxed. Note that both cases start relaxing at the same time and relax to the same maximum relaxation. However, case (b) reaches full relaxation earlier in the cycle when compared to case (a).

\begin{figure*}
    \centering{{\includegraphics[trim=10 200 30 200 ,clip,width=0.7\textwidth]{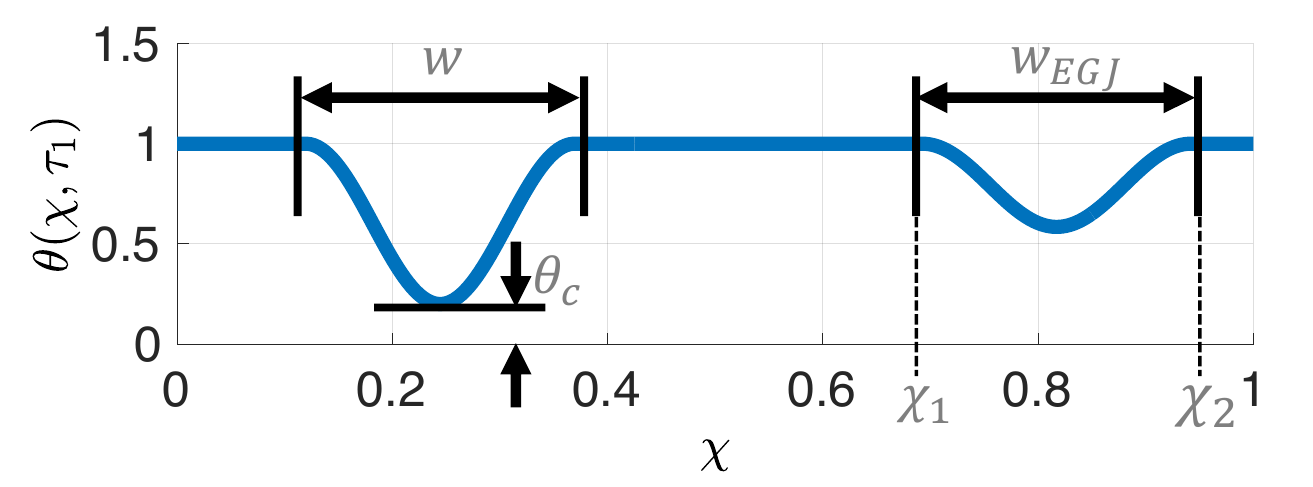}}}
    \caption{A plot of the activation function $\theta$ from equation (\ref{eq:thetaSuperposition}) along the tube length at a single time instance $\tau_1$.}
    \label{fig:theta_plot}
\end{figure*}

Figure \ref{fig:theta_plot} presents a plot of the activation function $\theta$ at time instance $\tau_1$. The figure shows how functions $\theta_p$ and $\theta_{\text{EGJ}}$ apply simultaneously to the tube wall. While the peristaltic contraction wave travels down the length of the tube at constant contraction ($\theta_c$) and speed ($c$), the tone at the EGJ relaxes.

\subsection{Boundary and Initial Conditions} \label{Boundary_Conditions}

The elastic tube is closed on both ends and the volume inside the tube remains constant such that there is no flow in or out of the tube \cite{Acharya_2021,Elisha2021}. Therefore 
\begin{equation} \label{eq:velBC}
    U\left(\chi=0,\tau\right)=0\qquad\text{and}\qquad U\left(\chi=1,\tau\right)=0.
\end{equation}

\noindent The boundary condition for $\alpha$ is obtained by applying the velocity boundary condition in equation (\ref{eq:velBC}) into equation (\ref{eq:momentum_nondim}), which implies that $\partial p/\partial \chi = 0$ \cite{Acharya_2021,Elisha2021}. Therefore, by taking the spatial derivative of equation (\ref{eq:tube_law_nondim}) and setting it equal to zero, we obtain a Neumann boundary condition for $\alpha$ (the non-dimensional cross-sectional area) of the form,

\begin{equation} \label{eq:areaBC}
    \left.\frac{\partial}{\partial \chi} \left(\frac{\alpha}{\theta}\right)\right|_{\chi=0,\tau}=0\qquad\mathrm{and}\qquad
    \left.\frac{\partial}{\partial \chi} \left(\frac{\alpha}{\theta}\right)\right|_{\chi=1,\tau}=0.
\end{equation}
Note that this condition is derived assuming that $\eta\approx 0$. However, as explained in \cite{Acharya_2021}, the effects due to this damping at the boundary is negligible even when $\eta$ is not equal to zero.

At $\tau=0$, the traveling peristaltic wave is yet to enter the domain and the EGJ is contracted, therefore, the fluid velocity inside the tube is equal to zero everywhere \cite{Acharya_2021}. The cross-sectional area at $\tau=0$ depends on the constant volume of the tube and the activation function $\theta(\chi,\tau=0)$. Hence, the initial conditions are specified as such: 

\begin{equation} \label{eq:velareaIC}
    U\left(\chi,\tau=0\right)=0  \qquad\qquad  \alpha\left(\chi,\tau=0\right)=\alpha_{\text{IC}}\theta(\chi,\tau=0),
\end{equation}
where $\alpha_{\text{IC}}$ is a constant.

\subsection{Numerical Implementation} \label{Numerical_Implementation}

The MATLAB $\tt{pdepe}$ function is used to obtain the numerical solution of $\alpha(\chi,\tau)$ and $U(\chi,\tau)$ by solving equations (\ref{eq:continuity_nondim}) and (\ref{eq:momentum_nondim_final}) with the boundary and initial conditions in equations (\ref{eq:velBC}), (\ref{eq:areaBC}), and (\ref{eq:velareaIC}), respectively. The simulations are differentiated by a unique combination of the physical parameters defining this problem, listed in table \ref{table:parameterList}. The computational procedure to solve this forward simulation is described in greater details by \cite{Acharya_2021}, who also validates the 1D model by comparing it to an equivalent 3D immersed boundary simulation. 


\begin{table}[h!]
\centering
\caption{List of non-dimensional physical parameters}
\begin{tabular}{|| l l l ||} 
 \hline
 Symbol & Values  & Definition \\ [0.5ex] 
 \hline\hline
 $\beta$ & $100-10,000$ & Dimensionless strength of viscous effects (inverse of Reynolds number)  \\ 
 $\psi$ & $100-10,000$ & Dimensionless rigidity of the elastic tube (inverse of Cauchy number)\\
 $\gamma$ & $0.06-0.6$ & Dimensionless EGJ relaxation speed  \\
 $\theta_c$ & $0.05-0.2$ & Peristaltic contraction strength  \\
 $\theta_r$ & $0.05-0.2$ & EGJ contraction strength  \\
 $\text{w}$ & $0.25$ & Width of peristaltic wave  \\
 $\text{w}_{\text{EGJ}}$ & $0.25$ & EGJ width  \\ [1ex]
 \hline
\end{tabular}
\label{table:parameterList}
\end{table}

\section{Work Distribution in the System} \label{Work_MathDetails}

As mentioned in section \ref{PressureAreaLoops_Intro}, the fact that the opening and closing curves on the pressure-area plot create a loop rather than take the same path indicates that there is some energy that is being gained or lost by the EGJ walls. Therefore, in this section, we wish to derive an equation to examine the way in which energy is spent during a contractile cycle, which helps us to reveal the leading parameters that control the loop type. A similar study by Acharya et al. looked at the energy distribution and the work done by a peristaltic wave along the entire tube length without EGJ contraction \cite{Acharya_2021}. In this work, we focus on the energy spent at the EGJ region.

To derive the work distribution equation, we multiply the momentum equation (\ref{eq:momentum}) by the flow rate $Q=Au$, and rewrite it such that

\begin{equation}\label{eq:momentumQ}
A\frac{\partial}{\partial t}\left(\frac{1}{2}\rho u^{2}\right)+ 
Q\frac{\partial}{\partial x}\left(\frac{1}{2}\rho u^{2}\right)=
-Au\frac{\partial P}{\partial x}-\frac{8\pi\mu uQ}{A}.
\end{equation}

\noindent Next, we add the continuity equation to the left hand side of equation (\ref{eq:momentumQ}) and rearrange it to obtain

\begin{equation}
\frac{\partial}{\partial t}\left(\frac{1}{2}\rho Au^{2}\right) +
\frac{\partial}{\partial x}\left(\frac{1}{2}\rho Qu^{2}\right) = 
-{\frac{\partial\left(AuP\right)}{\partial x}}
+{P\frac{\partial\left(Au\right)}{\partial x}}
-{\frac{8\pi\mu uQ}{A}.}
\end{equation}

\noindent From equation (\ref{eq:continuity}), we can replace $\partial\left(Au\right)/\partial x$ with $-\partial A/\partial t$, to acquire the final form of the momentum equation

\begin{equation} \label{eq:start_point}
\frac{\partial}{\partial t}\left(\frac{1}{2}\rho Au^{2}\right) + 
\frac{\partial}{\partial x}\left(\frac{1}{2}\rho Qu^{2}\right) = 
-\frac{\partial\left(AuP\right)}{\partial x}-P\frac{\partial A}{\partial t}-\frac{8\pi\mu uQ}{A}.
\end{equation}

Integrating equation (\ref{eq:start_point}) with respect to length results in a power conservation equation of the form

\begin{equation}  \label{eq:power_balance_no_split}
\underbrace{-\int\limits _{x_1}^{x_2}P\frac{\partial A}{\partial t}\mathrm{d}x}_{\text{I}}
=\underbrace{\frac{\partial}{\partial t}\int\limits _{x_1}^{x_2}\left(\frac{1}{2}\rho Au^{2}\right)\mathrm{d}x}_{\text{II}} +
\underbrace{\int\limits _{x_1}^{x_2}8\pi\mu u^{2}\mathrm{d}x}_{\text{III}} +
\underbrace{\left(AuP\right)\bigg|_{x_1}^{x_2}}_{\text{IV}} + 
\underbrace{\left(\frac{1}{2}\rho Au^3\right)\bigg|_{x_1}^{x_2}}_{\text{V}}.
\end{equation}

\noindent The equation above portrays the power balance equation. The left hand side term represents the rate of work that is done by the tube wall, and the right hand side is equal to the consumers of this spent power. Each term in equation (\ref{eq:power_balance_no_split}) is numbered, where (I) is the rate of work done by the tube wall on the fluid, (II) is the rate of change in kinetic energy of the fluid, (III) is the rate of energy loss due to viscous dissipation, (IV) is the rate of work done by pressure on the cross-section, and (V) is the rate of change of momentum.

Integrating equation (\ref{eq:power_balance_no_split}) with respect to time results in the the work distribution equation, such that

\begin{equation}  \label{eq:work_balance_no_split}
\begin{aligned}
{-\int\limits _{t_1}^{t_2}\int\limits _{x_1}^{x_2}P\frac{\partial A}{\partial t}\mathrm{d}x\mathrm{d}t} &= {\int\limits _{t_1}^{t_2}\frac{\partial}{\partial t}\int\limits _{x_1}^{x_2}\left(\frac{1}{2}\rho Au^{2}\right)\mathrm{d}x\mathrm{d}t} +
{\int\limits _{t_1}^{t_2}\int\limits _{x_1}^{x_2}8\pi\mu u^{2}\mathrm{d}x\mathrm{d}t}\\
&+{\int\limits _{t_1}^{t_2}\left(AuP\right)\bigg|_{x_1}^{x_2}\mathrm{d}t} + 
{\int\limits _{t_1}^{t_2}\left(\frac{1}{2}\rho Au^3\right)\bigg|_{x_1}^{x_2}\mathrm{d}t}.
 \end{aligned}
\end{equation}

\noindent This form shows in details how work is distributed in the system.  The non-dimensional form of equation (\ref{eq:work_balance_no_split}) is

\begin{equation}  \label{eq:work_balance_NonDim_no_split}
\begin{aligned}
{-\psi\int\limits _{\tau_1}^{\tau_2}\int\limits _{\chi_1}^{\chi_2}p\frac{\partial \alpha}{\partial \tau}\mathrm{d}\chi\mathrm{d}\tau} &={\int\limits _{\tau_1}^{\tau_2}\frac{\partial}{\partial \tau}\int\limits _{\chi_1}^{\chi_2}\left(\frac{1}{2}\alpha U^{2}\right)\mathrm{d}\chi\mathrm{d}\tau} +
{\beta\int\limits _{\tau_1}^{\tau_2}\int\limits _{\chi_1}^{\chi_2}U^{2}\mathrm{d}\chi\mathrm{d}\tau} \\
& +{\psi\int\limits _{\tau_1}^{\tau_2}\left(\alpha Up\right)\bigg|_{\chi_1}^{\chi_2}\mathrm{d}\tau} + 
{\int\limits _{\tau_1}^{\tau_2}\left(\frac{1}{2}\alpha U^3\right)\bigg|_{\chi_1}^{\chi_2}\mathrm{d}\tau}.
 \end{aligned}
\end{equation}

\begin{figure*}
    \centering{{\includegraphics[clip,width=0.7\textwidth]{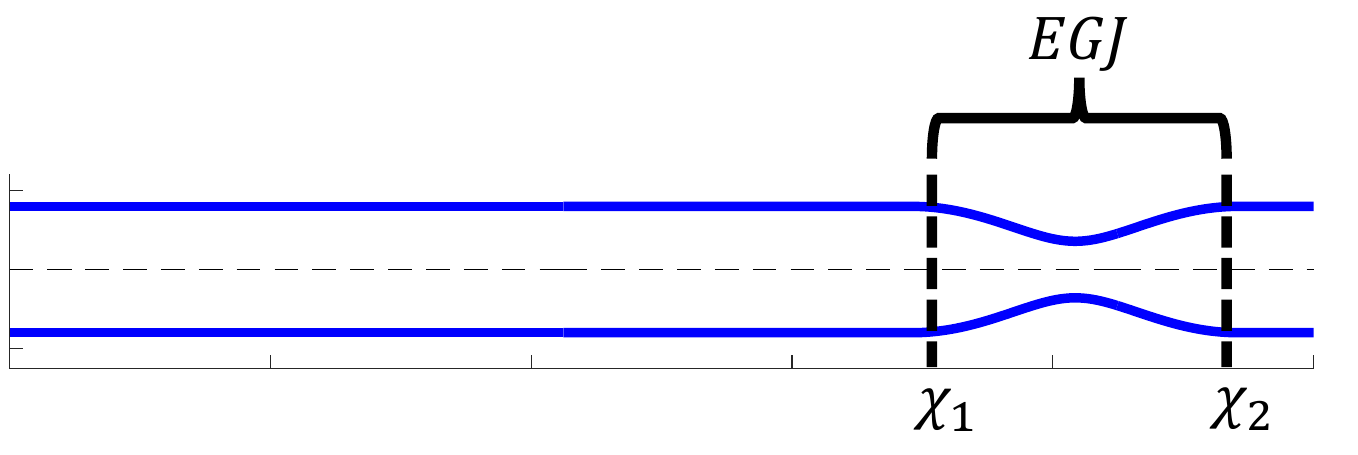}}}
    \caption{Image showing the EGJ location on the elastic tube at $\tau=0$.}
    \label{fig:EGJ_Location}
\end{figure*}

In our case, we want to look at the work distribution throughout the entire contractile cycle, and therefore, $\tau_1=0$ and $\tau_2=\tau_f$ where $\tau_f$ is the final, non-dimensional time instance. For work distribution along the entire esophagus length, we set $\chi_1=0$ and $\chi_2=1$. For work distribution at the EGJ, we specify an 'EGJ region' of  width $\text{w}_{\text{EGJ}}=\chi_2-\chi_1=0.25$ (seen in figure \ref{fig:EGJ_Location}), where $\chi_1=0.70$ and $\chi_2=0.95$.

\subsection{Passive \& Active Work Decomposition} \label{Passive_Active_Decomposition}
As noted by \cite{Acharya_2021}, the pressure term has contributions from both the passive expansion of the tube wall and the pressure rise or drop due to active contraction or relaxation of the wall \cite{Acharya_2021}. Hence, we decompose the total pressure into its active and passive components, such that

\begin{equation}
P(x,t)=P_{\text{passive}} + P_{\text{active}}.
\label{eqn:P_pass_act_sum}
\end{equation}

\noindent Therefore, the total work done by the tube wall on the fluid (LHS of equation (\ref{eq:work_balance_no_split})) is the sum of the passive and active work. This decomposition provides a deeper understanding into the way in which work progresses over time \cite{Acharya_2021}. Moreover, decomposing the total work into its active and passive components is an essential ingredient in explaining the difference between the two pressure-area loop types, as discussed in section \ref{results}.

The passive work is the elastic energy that is stored in the tube wall as the peristaltic contraction wave travels towards the distal end of the tube and increases the fluid pressure. It is the process by which the cross-sectional area changes while the activation function $\theta$ remains unchanged. The passive pressure is defined as 

\begin{equation}
P_{\text{passive}}(x,t)=K_e\left(\frac{A(x,t)}{A_{\scriptscriptstyle o}\theta_{\text{IC}}(x)}-1\right) + P_o
\label{eqn:PassiveDim}
\end{equation}

\noindent where $\theta_{\text{IC}}(x)=\theta(x,t=0)$. The non-dimensional form of equation (\ref{eqn:PassiveDim}) is
\begin{equation}
p_{\text{passive}}(\chi,\tau)=\frac{\alpha(\chi,\tau)}{\theta_{\text{IC}}(\chi)}-1.
\label{eqn:Passive}
\end{equation}

The active work is the work done by the active relaxation and contraction of the tube wall. It refers to the change in the tube's reference area through the activation function $\theta$ while keeping $A(x,t)$ fixed. The active pressure is defined as 

\begin{equation}
P_{\text{active}}(x,t)=K_e\frac{A(x,t)}{A_{\scriptscriptstyle o}}\left(\frac{1}{\theta(x,t)}-\frac{1}{\theta_{\text{IC}}(x)}\right)
\label{eqn:ActiveDim}
\end{equation}

\noindent and the non-dimensional form of the active pressure is
\begin{equation}
p_{\text{active}}(\chi,\tau)=\alpha\left(\frac{1}{\theta(\chi,\tau)}-\frac{1}{\theta_{\text{IC}}(\chi)}\right).
\label{eqn:Active}
\end{equation}

Using this decomposition, equation (\ref{eq:work_balance_no_split}) can be written as

\begin{equation}  \label{eq:work_balance_Dim}
\begin{aligned}
{-\int\limits _{t_1}^{t_2}\int\limits _{x_1}^{x_2}P_{\text{active}}\frac{\partial A}{\partial t}\mathrm{d}x\mathrm{d}t} &= {\int\limits _{t_1}^{t_2}\frac{\partial}{\partial t}\int\limits _{x_1}^{x_2}\left(\frac{1}{2}\rho Au^{2}\right)\mathrm{d}x\mathrm{d}t} +
{\int\limits _{t_1}^{t_2}\int\limits _{x_1}^{x_2}8\pi\mu u^{2}\mathrm{d}x\mathrm{d}t}\\
&+{\int\limits _{t_1}^{t_2}\left(AuP\right)\bigg|_{x_1}^{x_2}\mathrm{d}t} + 
{\int\limits _{t_1}^{t_2}\left(\frac{1}{2}\rho Au^3\right)\bigg|_{x_1}^{x_2}\mathrm{d}t} +
{\int\limits _{t_1}^{t_2}\int\limits _{x_1}^{x_2}P_{\text{passive}}\frac{\partial A}{\partial t}\mathrm{d}x\mathrm{d}t},
 \end{aligned}
\end{equation}
and equation (\ref{eq:work_balance_NonDim_no_split}) becomes

\begin{equation}  \label{eq:work_balance}
\begin{aligned}
{-\psi\int\limits _{\tau_1}^{\tau_2}\int\limits _{\chi_1}^{\chi_2}p_{\text{active}}\frac{\partial \alpha}{\partial \tau}\mathrm{d}\chi\mathrm{d}\tau} &={\int\limits _{\tau_1}^{\tau_2}\frac{\partial}{\partial \tau}\int\limits _{\chi_1}^{\chi_2}\left(\frac{1}{2}\alpha U^{2}\right)\mathrm{d}\chi\mathrm{d}\tau} +
{\beta\int\limits _{\tau_1}^{\tau_2}\int\limits _{\chi_1}^{\chi_2}U^{2}\mathrm{d}\chi\mathrm{d}\tau} \\
& +{\psi\int\limits _{\tau_1}^{\tau_2}\left(\alpha Up\right)\bigg|_{\chi_1}^{\chi_2}\mathrm{d}\tau} + 
{\int\limits _{\tau_1}^{\tau_2}\left(\frac{1}{2}\alpha U^3\right)\bigg|_{\chi_1}^{\chi_2}\mathrm{d}\tau} +
{\psi\int\limits _{\tau_1}^{\tau_2}\int\limits _{\chi_1}^{\chi_2}p_{\text{passive}}\frac{\partial \alpha}{\partial \tau}\mathrm{d}\chi\mathrm{d}\tau},
 \end{aligned}
\end{equation}

\noindent which is the non-dimensional form of equation (\ref{eq:work_balance_Dim}). Notice that the passive term is on the RHS of the equation. In this form, the work done by the active part of the tube wall is equal to the consumers of this work, which includes the elastic energy stored in the tube wall.

It is important to realize that the proposed definitions for the passive and active pressure terms in equations (\ref{eqn:Passive}) and (\ref{eqn:Active}), respectively, are not random, as they maintain two profound physical conditions that must hold. First, note that at $\tau=0$, the traveling peristaltic wave is yet to enter the domain and the EGJ is contracted, therefore the active pressure must be equal to zero for all $\chi$. To verify that, recall that $\theta = \theta_{\text{IC}}$ at $\tau=0$. Plugging this into equation (\ref{eqn:Active}) yields $p_{\text{active}}=0$ for all $\chi$. Second, at $\tau=0$, the total pressure along the tube is uniform. To show that, recall that $\alpha(\chi,\tau=0)=\alpha_{\text{IC}}\theta_{\text{IC}}(\chi)$, where $\alpha_{\text{IC}}$ is a constant. Hence, using equation (\ref{eqn:Passive}) and the fact that $p_{\text{active}}=0$ at $\tau=0$, $p_{\text{passive}}=\alpha_{\text{IC}}-1=C$ for all $\chi$ where $C$ is a constant.

\section{Results and Discussion} \label{results}

In this section, we identify the key characteristics of each loop type and the physical parameters that cause one loop to occur over the other. This is done through looking at how energy is being distributed in the tube over time. In the first part of this section, we analyze the simulation results over the entire tube length but our main efforts focus on the EGJ section. Later on, we utilize the conclusions obtained by the simulation results and apply them to clinical data. 

\subsection{Work Curves of Entire Tube Length} \label{Work_EntireLength}

We first look at the work distribution in the system over time along the entire esophagus length. Since the tube is closed on both ends, the velocity at both ends is zero, and equation (\ref{eq:work_balance}) simplifies to

\begin{equation}  \label{eq:work_balanceFull}
\begin{aligned}
{-\psi\int\limits _{\tau_1}^{\tau_2}\int\limits _{\chi_1}^{\chi_2}p_{\text{active}}\frac{\partial \alpha}{\partial \tau}\mathrm{d}\chi\mathrm{d}\tau} &={\int\limits _{\tau_1}^{\tau_2}\frac{\partial}{\partial \tau}\int\limits _{\chi_1}^{\chi_2}\left(\frac{1}{2}\alpha U^{2}\right)\mathrm{d}\chi\mathrm{d}\tau} +
{\beta\int\limits _{\tau_1}^{\tau_2}\int\limits _{\chi_1}^{\chi_2}U^{2}\mathrm{d}\chi\mathrm{d}\tau} + {\psi\int\limits _{\tau_1}^{\tau_2}\int\limits _{\chi_1}^{\chi_2}p_{\text{passive}}\frac{\partial \alpha}{\partial \tau}\mathrm{d}\chi\mathrm{d}\tau}.
 \end{aligned}
\end{equation}

\begin{figure*}[!htb]
    \centering{{\includegraphics[trim=20 160 30 180,clip,width=0.5\textwidth]{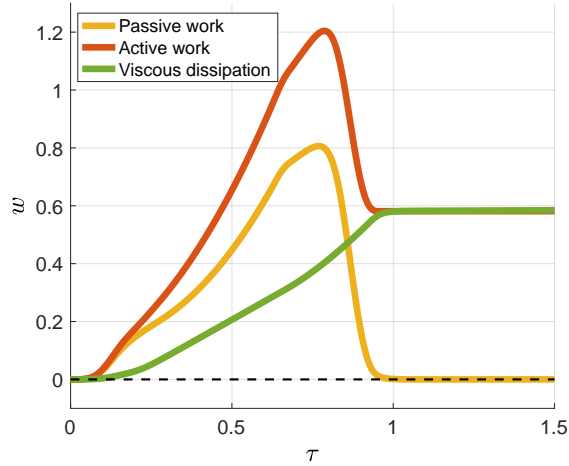}}}
    \caption{Passive, active, and viscous work curves along the entire esophagus length as a function of time. At each time instance, the sum of the energy loss dues to viscous dissipation and the passive energy stored in the tube wall is equal to the active work done by the tube wall.}
    \label{fig:workCurve_EntireLength}
\end{figure*}

\noindent Figure \ref{fig:workCurve_EntireLength} presents a plot of the work components in equation (\ref{eq:work_balanceFull}), normalized by $\psi$ ($w=W/\psi$), as a function of time. The data plotted in the figure are extracted from one simulation of a single contractile cycle, and the spatial integration is over the entire tube length. Note that the values of $\beta$ and $\psi$ considered in this analysis are much greater than 1, and therefore the kinetic energy term is very small compared to the other terms and thus not plotted.

As figure \ref{fig:workCurve_EntireLength} shows, at $\tau=0$, right before the peristaltic wave starts traveling into the domain, all the work components are equal to zero. However, once the contraction wave starts traveling and the EGJ begins relaxing ($\tau\approx0.1$), all three components start increasing, each having a unique pattern. Starting with the passive work, the elastic energy stored in the tube wall increases as a result of passive expansion of the tube wall, caused by pressure increase due to the traveling wave. This increase continues until $\tau\approx0.70$, when the peristaltic wave begins contracting the EGJ, making the tube cross sectional area closer to its original shape. Hence the passive energy starts decreasing. The sharp decrease continues until the end of the contractile cycle, when the passive energy recovers ($w_{\text{passive,final}}=0$), indicating that the tube returned to its original shape.

The active work curve has a similar pattern. It increases with time as a result of both the contraction wave and the EGJ relaxation. Once $\tau\approx0.70$, the active work that is defined by the change in $\theta$ starts decreasing since the contraction of the EGJ brings $\theta$ closer to $\theta_{\text{IC}}$. Lastly, the energy loss due to viscous dissipation curve has a different pattern. It maintains a steady increase throughout the contractile cycle, which ends when the peristaltic wave stops traveling. Since viscous dissipation cannot be recovered, $w_{\text{viscous}}\neq0$ at the end of the cycle. Hence, the active work at the end of the cycle is the energy that is being dissipated.

\subsubsection{Work Curves of EGJ Region} \label{Work_EGJ}

In this section, we only focus on a portion of the tube length, which we referred to as `EGJ region'. We determine the total work done by the EGJ wall over time and examine how this work relates to the pressure-area loop and the shape of the tube. Moreover, we look at the work distribution at the EGJ region over time by plotting the curves of each of the work components from equation (\ref{eq:work_balance}) as a function of time. The limits of the spatial integration in this case were discussed in section \ref{Work_MathDetails}.

\begin{figure*}
    \centering{{\includegraphics[trim=0 0 0 0,clip,width=0.8\textwidth]{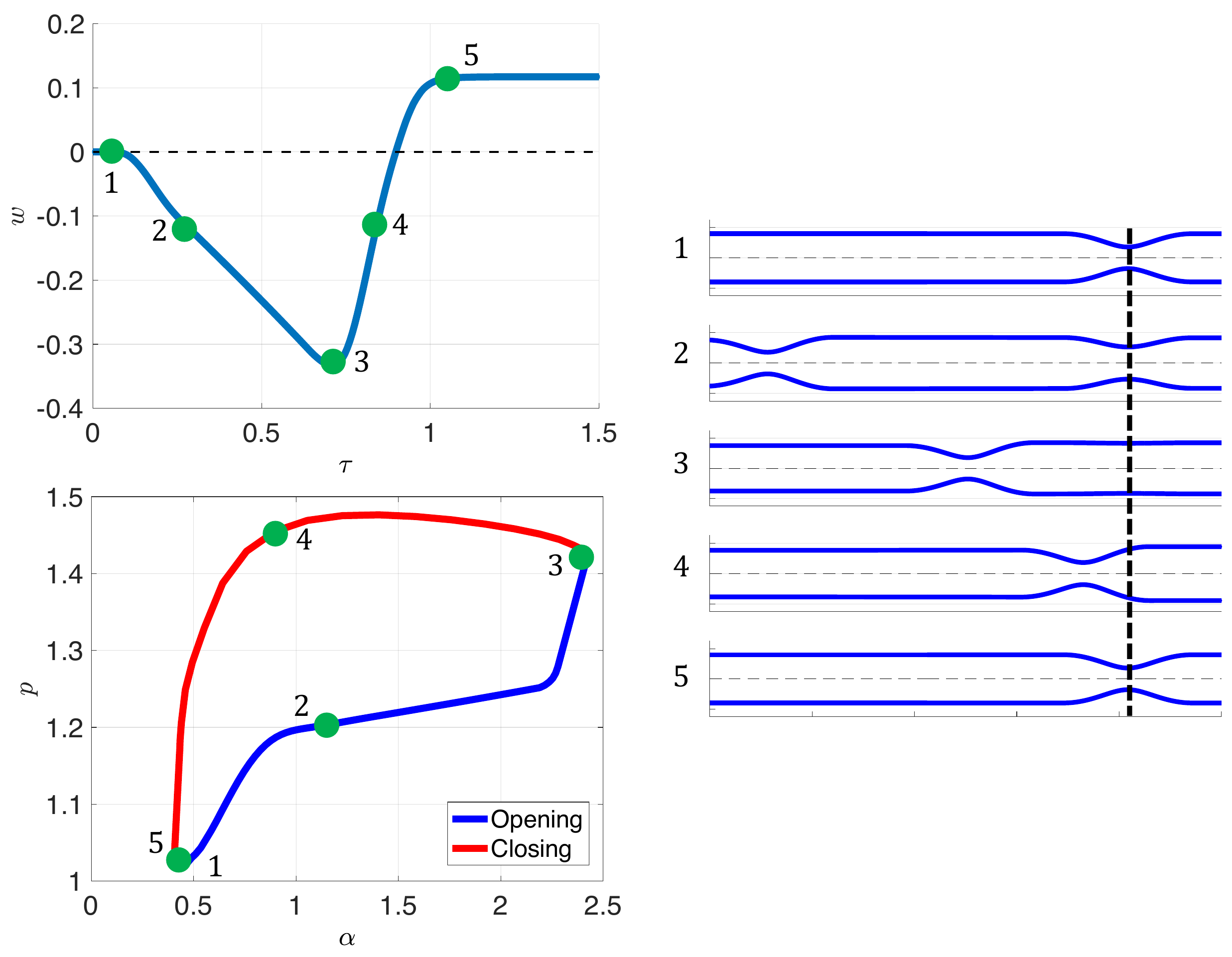}}}
    \caption{Visualizing simulation results and tracking work curve, pressure-area loop at EGJ location, and tube shape at five time instances. The graph at the top left shows the curve of total work done by the EGJ wall over time. The plot at the bottom left presents the corresponding pressure-area loop as recorded at the EGJ location. The right figure displays the tube shape at five different time instances, ordered chronologically, corresponding to the five points.}
    \label{fig:trackTime}
\end{figure*}

Figure \ref{fig:trackTime} presents the results of a single contractile cycle simulation. The graph at the top left displays the total work done by the EGJ wall as a function of time. The bottom left figure presents the corresponding pressure-area loop at the EGJ location. The five points highlighted on the loop and the work curve represent five time instances in the contractile cycle. The tube shape at these five instances are plotted on the right. Highlighting these instances in time allows us to relate between the total work done by the EGJ wall and the pressure-area loop, which helps us to better understand the elements that characterize the loops.

\begin{figure*}
    \centering{{\includegraphics[trim=20 120 30 130,clip,width=0.6\textwidth]{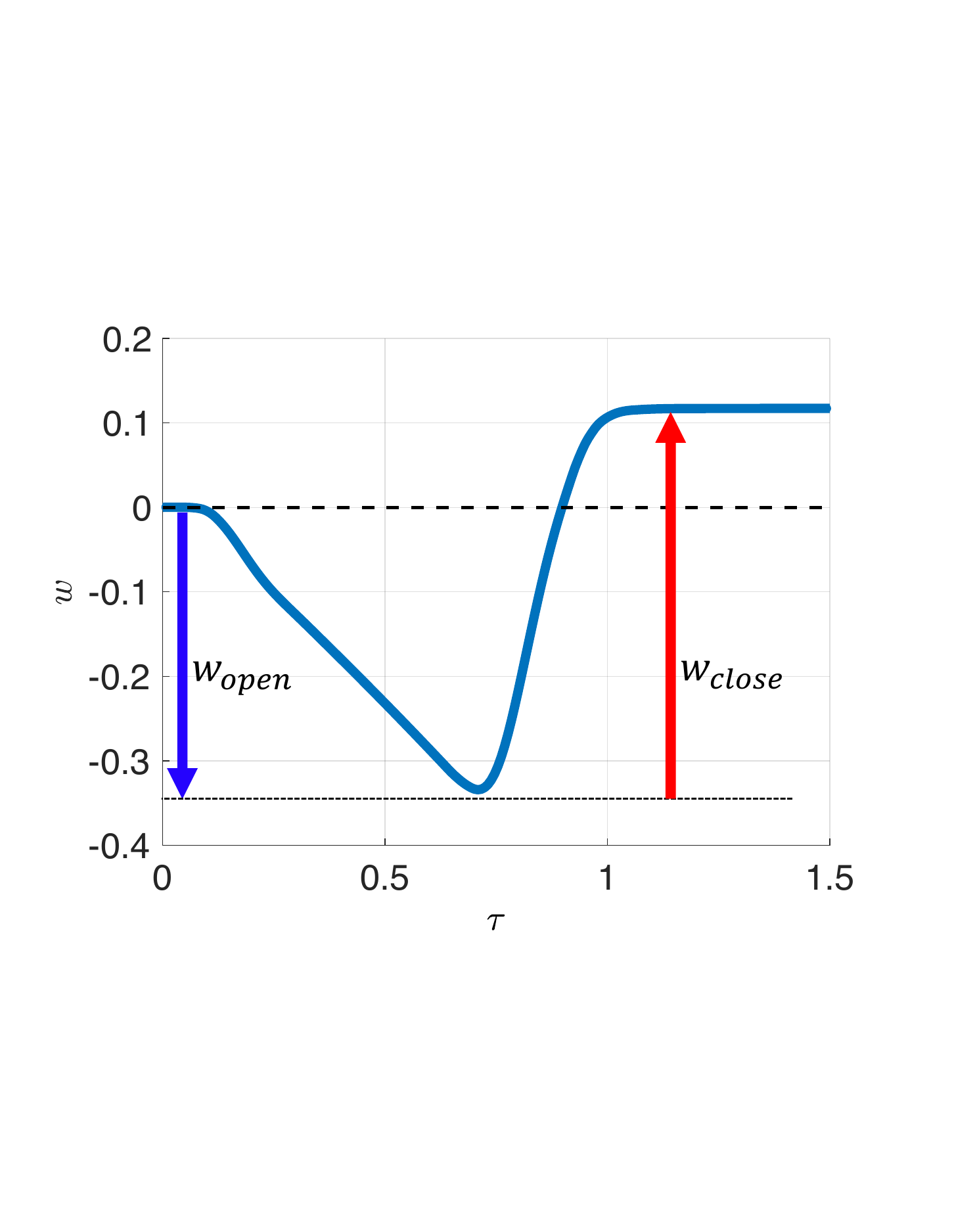}}}
    \caption{Graph of the total work done by the EGJ wall during opening and closing, as a function of time.}
    \label{fig:open_close_work_EGJ}
\end{figure*}

\begin{description}
  \item[\boldmath${\tau\leq\tau_1}$.] At the first time instance, right before the peristaltic contraction enters the domain, the EGJ is contracted and the total work done by the EGJ wall is equal to zero. 
  
  \item[\boldmath${\tau_1<\tau<\tau_3}$.] Between $\tau_1$ and $\tau_3$, the peristaltic contraction wave travels along the length of the tube and the EGJ actively relaxes simultaneously. Together, they result in the increase of the cross-sectional area and the pressure at the EGJ, as seen in the loop in figure \ref{fig:trackTime}. Since the EGJ wall opens, it does negative work on the fluid, which explains why the work curve at the top left of figure \ref{fig:trackTime} goes negative between $\tau_1$ and $\tau_3$. The total work done by the EGJ wall during opening is therefore $w_{\text{open}}={-\psi\int\limits _{\tau_1}^{\tau_3}\int\limits _{\chi_1}^{\chi_2}p\frac{\partial \alpha}{\partial \tau}\mathrm{d}\chi\mathrm{d}\tau}$, represented by the blue arrow in figure \ref{fig:open_close_work_EGJ}.
  
  \item[\boldmath${\tau=\tau_3}$.] At the third time instance, the EGJ is fully open, corresponding to the minimum point on the work plot in figure \ref{fig:trackTime}.
  
  \item[\boldmath${\tau_3<\tau<\tau_5}$.] Between $\tau_3$ and $\tau_5$, the peristaltic wave travels into the EGJ region which eventually closes the EGJ. In this process, the cross-sectional area and pressure at the EGJ generally decrease as seen in the loop in figure \ref{fig:trackTime}. Since the EGJ wall closes, it squeezes the fluid, applying positive work, which is seen on the work plot in \ref{fig:trackTime} by an increase in work between points 3 and 5. The total work done by the EGJ wall during closing is therefore $w_{\text{close}}={-\psi\int\limits _{\tau_3}^{\tau_5}\int\limits _{\chi_1}^{\chi_2}p\frac{\partial \alpha}{\partial \tau}\mathrm{d}\chi\mathrm{d}\tau}$, represented by the red arrow in figure \ref{fig:open_close_work_EGJ}.
  
   \item[\boldmath${\tau\geq\tau_5}$.]At the fifth time instance, the peristaltic wave stops traveling and the tube returns to its initial shape. The work value at this instance is equivalent to the work done by the EGJ wall during the entire cycle.
\end{description}

The work done by the EGJ wall throughout the entire contractile cycle is equal to the difference in the magnitude of the work needed to close the EGJ to the work needed to open the EGJ. In figure \ref{fig:open_close_work_EGJ}, it is the difference between the length of the red and blue arrows. If the final value is positive, as in figure \ref{fig:trackTime}, more work is needed to close the EGJ than to open it. Moreover, it implies that in this contractile cycle, the EGJ wall does more work on the fluid than the fluid pressure does on the wall. On the other hand, if the work value at the end of the cycle is negative, more work is needed to open the EGJ than to close it, and fluid pressure does more work on the EGJ wall than the EGJ wall applies on the fluid throughout this contractile cycle. From our simulation results, we observed that the work done by the EGJ wall throughout the entire contractile cycle is negative for pressure dominant loops (figure \ref{fig:PDL_clinical}) and positive for tone dominant loops (figure \ref{fig:TDL_clinical}).

\begin{figure*}
    \centering{{\includegraphics[trim=17 95 20 105,clip,width=0.6\textwidth]{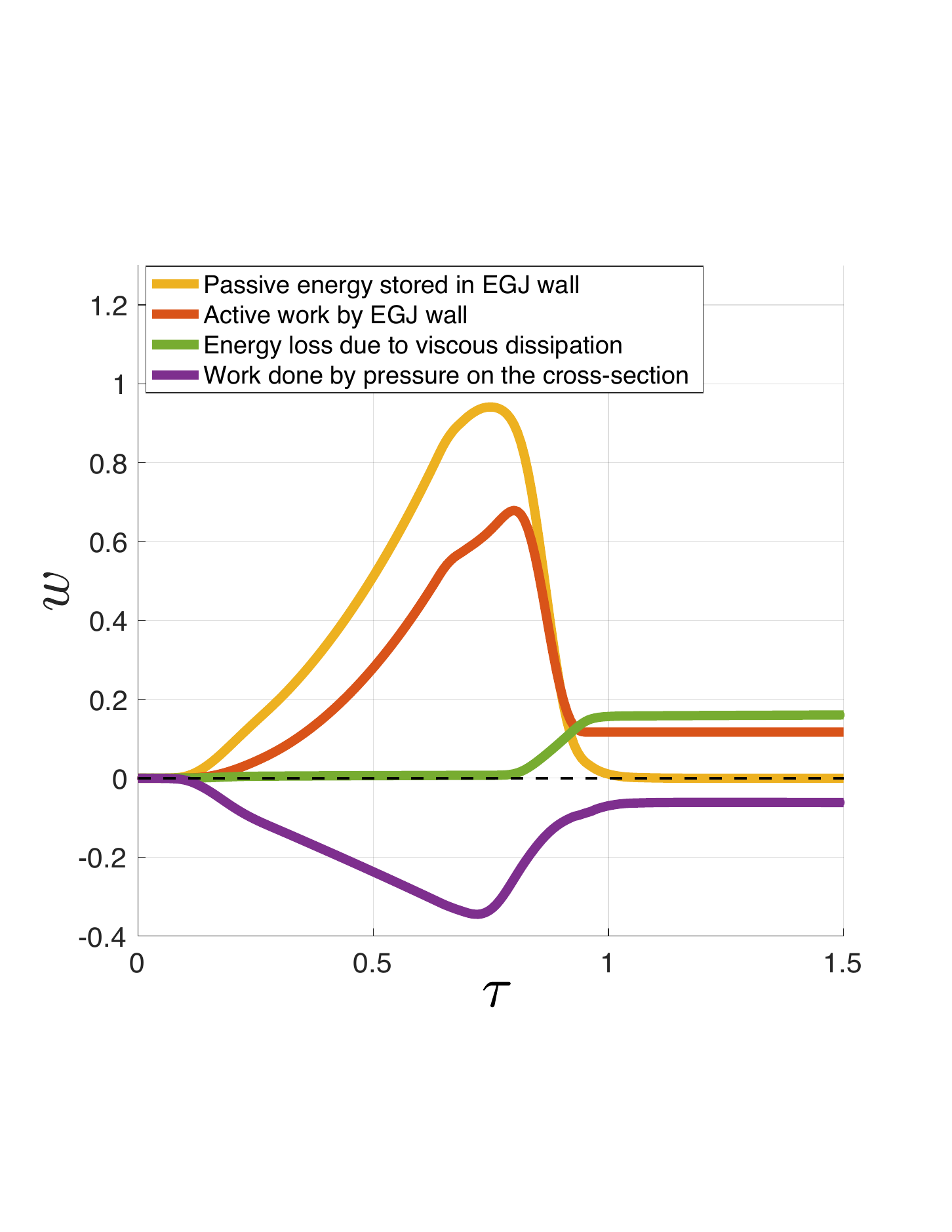}}}
    \caption{Plot of the work components from equation (\ref{eq:work_balance}) at the EGJ region as they progress over time.}
    \label{fig:workCurve_EGJ}
\end{figure*}

To explain this trend, we need to understand what makes $|w_{\text{close}}|>|w_{\text{open}}|$ a tone dominant loop as opposed to when $|w_{\text{close}}|<|w_{\text{open}}|$ a pressure dominant loop. Figure \ref{fig:workCurve_EGJ} presents the work components from equation (\ref{eq:work_balance}) evaluated at the EGJ region as a function of time. The kinetic energy and net rate of change of momentum terms are not plotted in the figure since $\psi>>1$ and $\beta>>1$. As the figure shows, the active, passive, and viscous dissipation terms display similar trends to the ones described in section \ref{Work_EntireLength} with two exceptions. First, the total energy that is being dissipated at the end of the contractile cycle is overcame by both the active work and the work done by pressure on the cross-section. Second, during EGJ opening, the leading source of energy is the passive expansion, where $|w_{\text{open,passive}}|>|w_{\text{open,active}}|$. 

Hence, the opening of the EGJ is led by the peristaltic contraction wave which increases the pressure and cross-sectional area at the EGJ. On the other hand, the closing of the EGJ is led by the active contraction of the EGJ tone, which needs to overcome the fluid that has accumulated at the EGJ location. Therefore, if $|w_{\text{close}}|>|w_{\text{open}}|$, the leading source of energy in this cycle is the EGJ tone, thus the name tone dominant loop. However, if $|w_{\text{close}}|<|w_{\text{open}}|$, the leading source of energy in this contractile cycle at the EGJ is the increase in fluid pressure due to the peristaltic wave, which opens the EGJ. Hence, the name pressure dominant loop.

Notice that although the active work at the EGJ represents relaxation (active opening of the EGJ contraction), its value is positive in figure \ref{fig:workCurve_EGJ}, and it increases as the EGJ contraction opens. This means that the EGJ wall does positive active work on the fluid during the opening. This occurs because when relaxation starts, the active pressure, as defined in equation (\ref{eqn:Active}) is negative, which implies that the fluid is sucking the EGJ wall down. Therefore, as relaxation occurs, the reference area at the EGJ starts growing, but because of the suction, the actual cross-sectional area of the EGJ is smaller than the reference area, so the wall pushes itself to a new relaxed location. The active work is the work done by the EGJ wall to match the actual cross-sectional area to the reference area.

\subsection{Key Parameters in Loop Type} \label{Parameters_Loop_Type}

Using the simulation results, we conducted a parametric study to examine the leading parameters that dictate the loop type. Out of the seven parameters listed in table \ref{table:parameterList}, only three parameters affect the loop type. These parameters are $\psi$, $\beta$, and $\gamma$. The rest of the parameters affect the loop quantitatively, such as slightly changing the loop shape or area, but do not contribute to the conversion between the loop types. In this section, we discuss the affect of $\psi$, $\beta$, and $\gamma$ on the loop type through looking at the total, passive and active work done by the EGJ wall.

\subsubsection{Tube Stiffness} \label{Tube_Stiffness}

\begin{figure*}
    \centering
    \begin{subfigure}[b]{0.48\textwidth}
        \centering
        {\includegraphics[trim=20 190 40 190,clip,width=\textwidth]{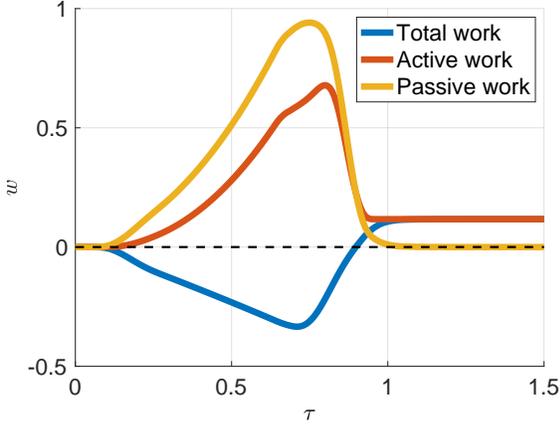}}
        \caption{Work curves by EGJ wall for $\psi = 100$}
        \label{fig:LowStiffWork}
    \end{subfigure}
    \hfill
    \begin{subfigure}[b]{0.48\textwidth}  
        \centering 
        {\includegraphics[trim=20 190 40 190,clip,width=\textwidth]{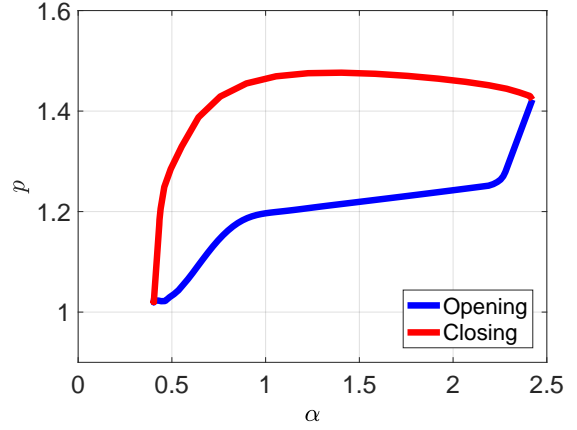}}
        \caption{Pressure-area loop at EGJ for $\psi = 100$}
        \label{fig:LowStiffLoop}
    \end{subfigure}
     \hfill
         \begin{subfigure}[b]{0.48\textwidth}  
        \centering 
        {\includegraphics[trim=20 190 40 190,clip,width=\textwidth]{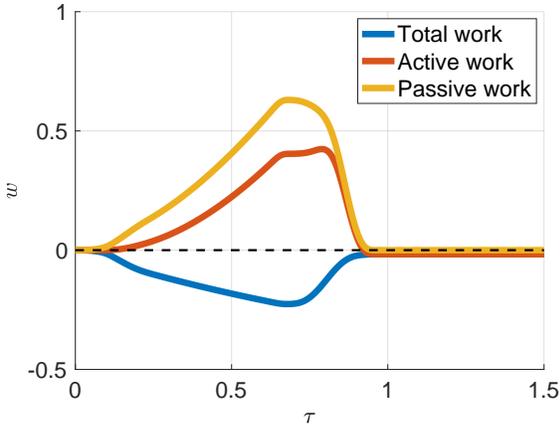}}
        \caption{Work curves by EGJ wall for $\psi = 2,400$}
        \label{fig:HighStiffWork}
    \end{subfigure}
     \hfill
    \begin{subfigure}[b]{0.48\textwidth}  
        \centering 
        {\includegraphics[trim=20 190 40 190,clip,width=\textwidth]{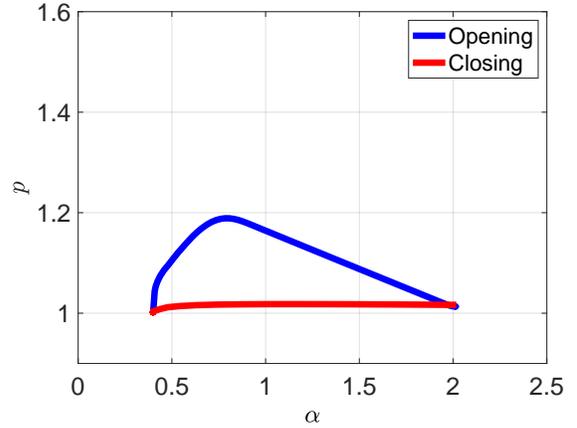}}
        \caption{Pressure-area loop at EGJ for $\psi = 2,400$}
        \label{fig:HighStiffLoop}
    \end{subfigure}
    \caption{Comparing two simulations with different $\psi$ values.} 
    \label{fig:wall_stiffness_comparison}
\end{figure*}

Figure \ref{fig:wall_stiffness_comparison} presents the work plots and pressure area loops of two different contractile cycle simulations. In these two simulations, all the parameters are the same but the value of $\psi$. Figures \ref{fig:LowStiffWork} and \ref{fig:LowStiffLoop} display the work curves and pressure-area loop for the simulation where $\psi=100$, respectively. Figures \ref{fig:HighStiffWork} and \ref{fig:HighStiffLoop} display the work curves and pressure-area loop for the simulation where $\psi=2,400$, respectively. As the figure shows, change in $\psi$ is sufficient to affect the loop type, where high values of $\psi$ are associated with pressure dominant loops (figure \ref{fig:PDL_clinical}) and low values of $\psi$ are associated with tone dominant loops (figure \ref{fig:TDL_clinical}). Note that the parameter $\psi=K_e/\rho c^2$ is dominated by the tube stiffness ($K_e$), and therefore, increase in $\psi$ can be looked at as increase in tube stiffness. Hence, pressure dominant loop is associated with a stiff esophagus and tone dominant loop is associated with a more compliment esophagus.

The lower the stiffness, the more compliant the tube is, and the distal cross-sectional area expands further.  The extended expansion of the distal tube wall means that a large amount of fluid has accumulated at the EGJ location, which increases fluid resistance \cite{Elisha2021}.  Hence, when the EGJ starts closing, the resistance is high, and the EGJ tone needs to apply more work in order to close. Consequently, we see that $|w_{\text{close}}|>|w_{\text{open}}|$ and a corresponding tone dominant loop. In the high stiffness tube, the increase in fluid pressure due to the peristaltic wave does not lead to a significant increase in the cross-sectional area distal of contraction (at the EGJ location). Therefore, the fluid is more equally distributed along the tube length and less work is required to contract the EGJ wall than to expand it. Hence, when stiffness is high, $|w_{\text{close}}|<|w_{\text{open}}|$ and the corresponding loop is a pressure dominant loop.

\subsubsection{Fluid Viscosity} \label{Fluid_Viscosity}

\begin{figure*}
    \centering
    \begin{subfigure}[b]{0.48\textwidth}
        \centering
        {\includegraphics[trim=20 190 40 190,clip,width=\textwidth]{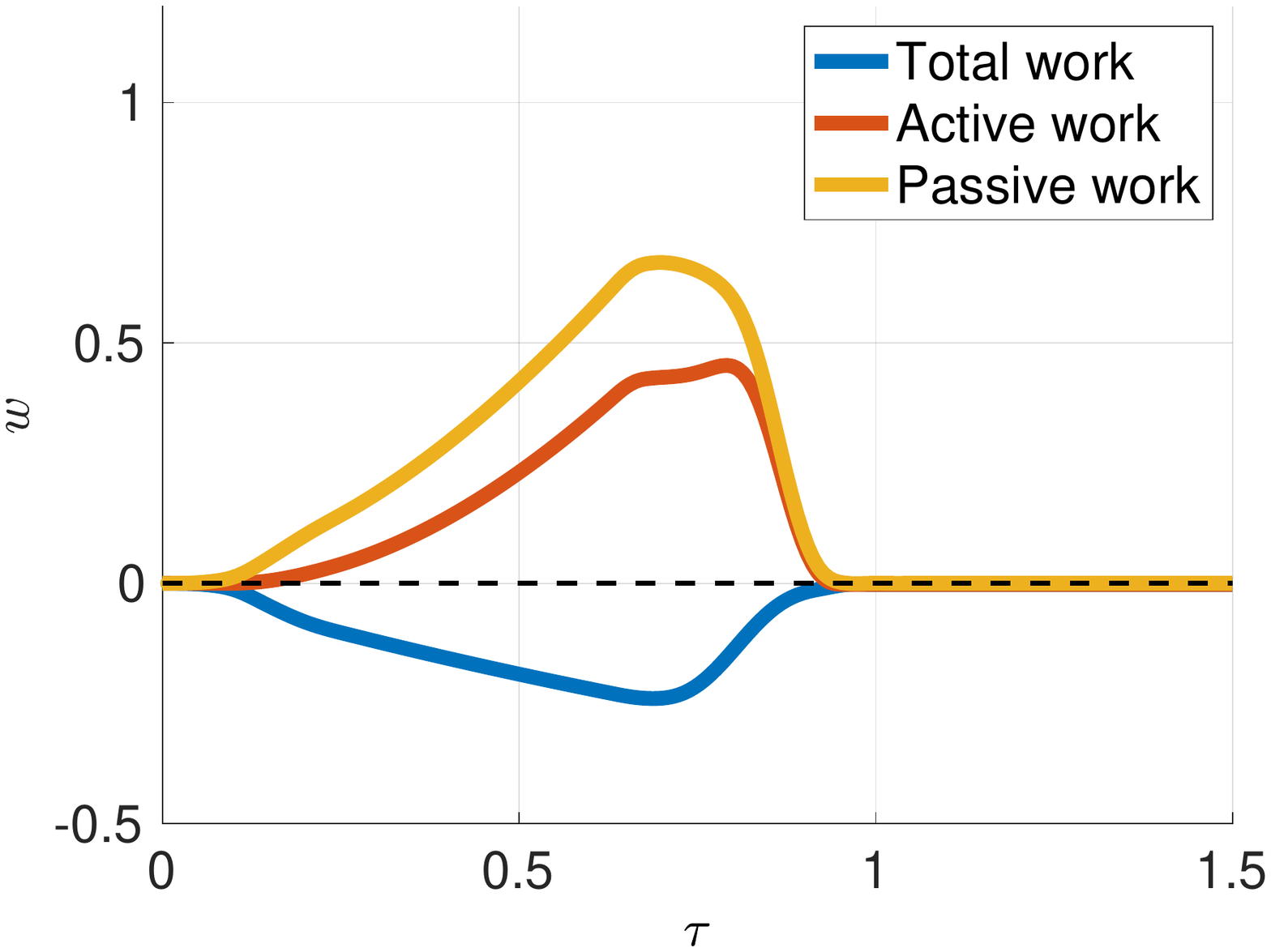}}
        \caption{Work curves by EGJ wall for $\beta = 100$}
        \label{fig:LowViscWork}
    \end{subfigure}
    \hfill
    \begin{subfigure}[b]{0.48\textwidth}  
        \centering 
        {\includegraphics[trim=20 190 40 190,clip,width=\textwidth]{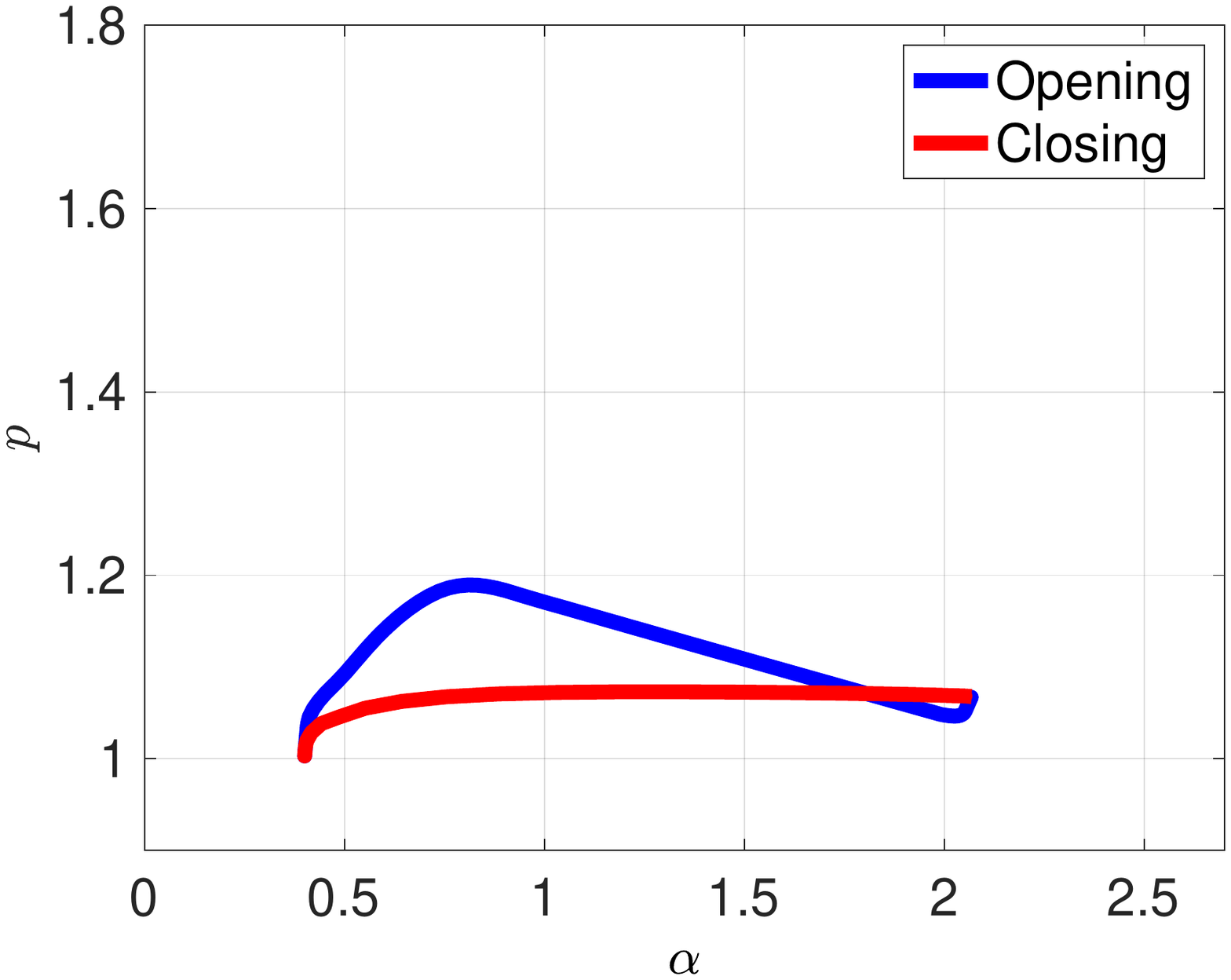}}
        \caption{Pressure-area loop at EGJ for $\beta = 100$}
        \label{fig:LowViscLoop}
    \end{subfigure}
     \hfill
         \begin{subfigure}[b]{0.48\textwidth}  
        \centering 
        {\includegraphics[trim=20 190 40 190,clip,width=\textwidth]{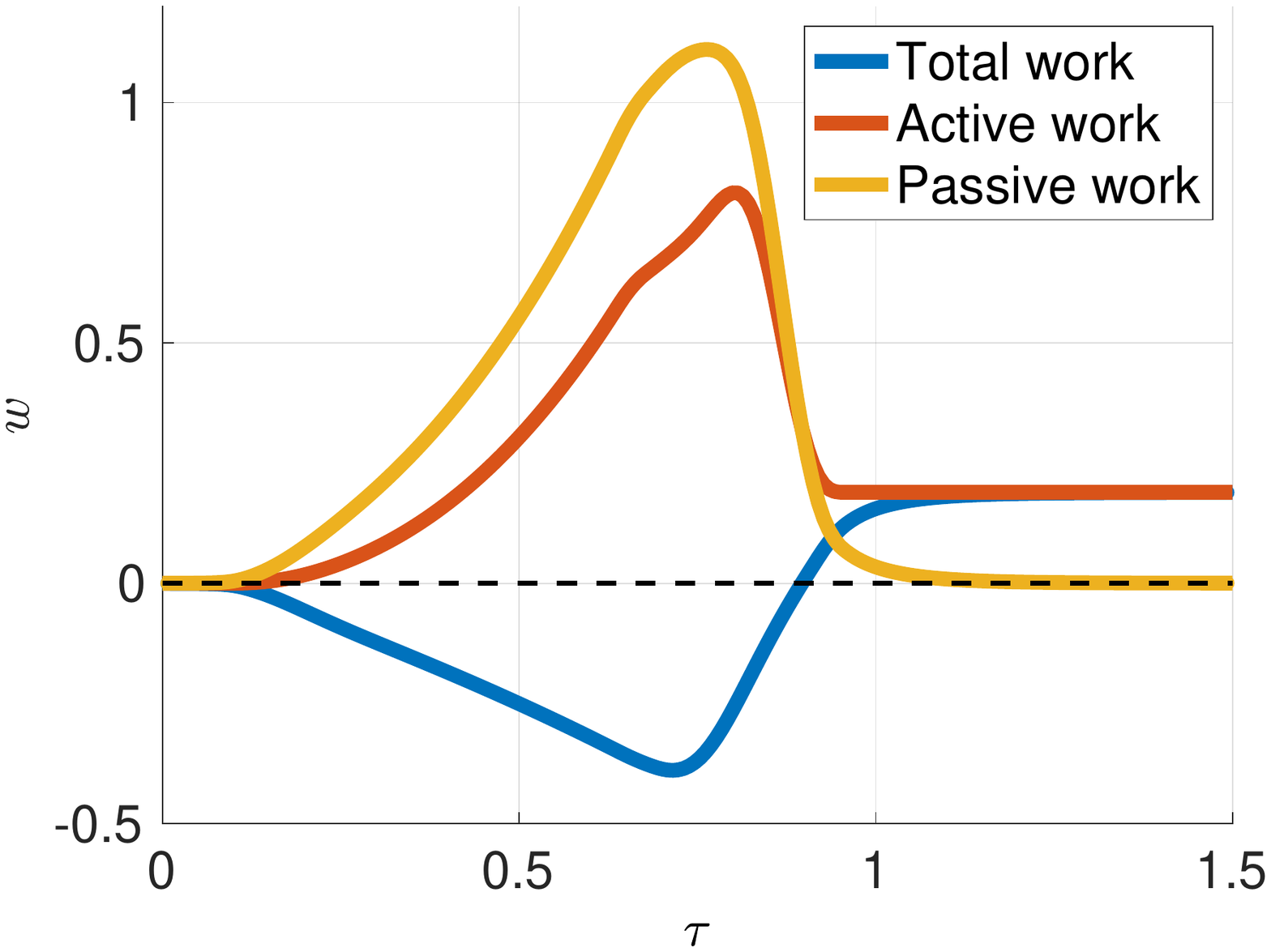}}
        \caption{Work curves by EGJ wall for $\beta = 1,000$}
        \label{fig:HighViscWork}
    \end{subfigure}
     \hfill
    \begin{subfigure}[b]{0.48\textwidth}  
        \centering 
        {\includegraphics[trim=20 190 40 190,clip,width=\textwidth]{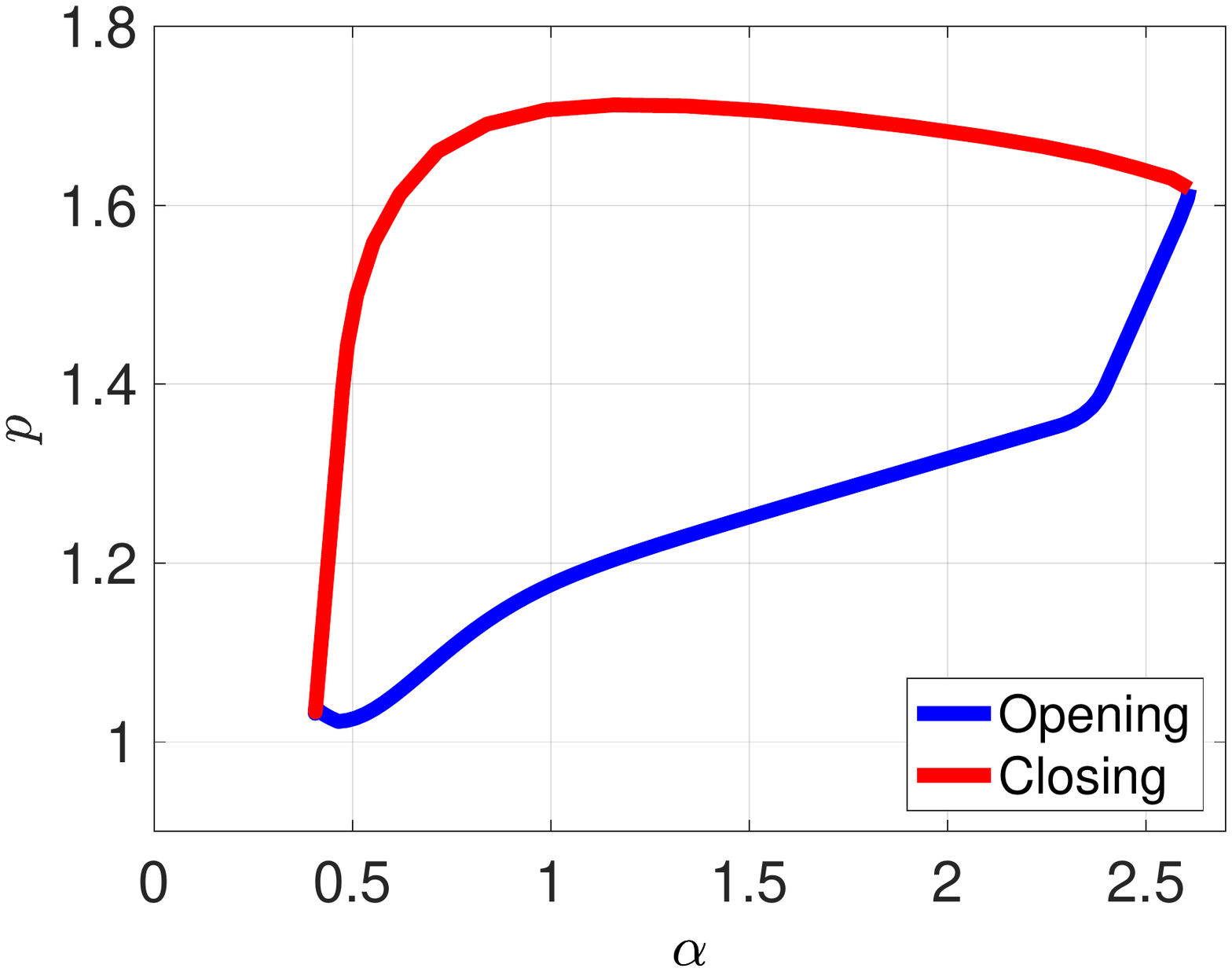}}
        \caption{Pressure-area loop at EGJ for $\beta = 1,000$}
        \label{fig:HighViscLoop}
    \end{subfigure}
    \caption{Comparing two simulations with different $\beta$ values.} 
    \label{fig:fluid_viscosity_comparison}
\end{figure*}

We can do a similar analysis as in section \ref{Tube_Stiffness}, but this time, we focus on parameter $\beta=8\pi\mu L/\rho A_oc$. Figure \ref{fig:fluid_viscosity_comparison} presents the work plots and pressure area loops of two different contractile cycle simulations of the same input parameters except for the value of $\beta$. In the work plot and loop in figures \ref{fig:LowViscWork} and \ref{fig:LowViscLoop}, respectively, $\beta=100$, and in the work plot and loop in figures \ref{fig:HighViscWork} and \ref{fig:HighViscLoop}, respectively, $\beta=1,000$. As the figure shows, change in $\beta$ is sufficient to affect the loop type, where low values of $\beta$ are associated with pressure dominant loops (figure \ref{fig:PDL_clinical}) and high values of $\beta$ are associated with tone dominant loops (figure \ref{fig:TDL_clinical}). Note that the parameter $\beta$ is dominated by the fluid viscosity ($\mu$), and therefore, increasing in $\beta$ can be looked at as increase in fluid viscosity. Hence, pressure dominant loop is associated with low fluid viscosity and tone dominant loop is associated with high fluid viscosity.

In order to explain this observation, we refer again to the work curves. As concluded in \cite{Elisha2021}, as long as the traveling peristaltic wave remains contracted, the fluid resistance increases with the increase in fluid viscosity. The increase in fluid resistance causes an overall higher pressure buildup distal of contraction, which helps to expand the cross-sectional area at the EGJ region (open the EGJ) \cite{Elisha2021}. Therefore, $|w_{\text{open,}\beta=100}|<|w_{\text{open,}\beta=1,000}|$. However, the opening work alone does not determine the loop type. The higher the pressure buildup at the EGJ when fluid viscosity is high implies that the EGJ wall needs to apply more work on the fluid in order contract, such that $|w_{\text{close}}|>|w_{\text{open}}|$. Hence, when fluid viscosity is high, the leading source of energy in this cycle is the EGJ tone, and a tone dominant loop emerges. On the other hand, when the viscosity is low and the resistance is low, the pressure at the distal end of the tube is lower, and the EGJ tone does not need to apply as much work in order to contract back. Hence, $|w_{\text{close}}|<|w_{\text{open}}|$ and a pressure dominant loop emerges.

\subsubsection{EGJ Relaxation Speed} \label{Opening_Speed}

\begin{figure*}
    \centering
    \begin{subfigure}[b]{0.48\textwidth}
        \centering
        {\includegraphics[trim=20 130 40 140,clip,width=\textwidth]{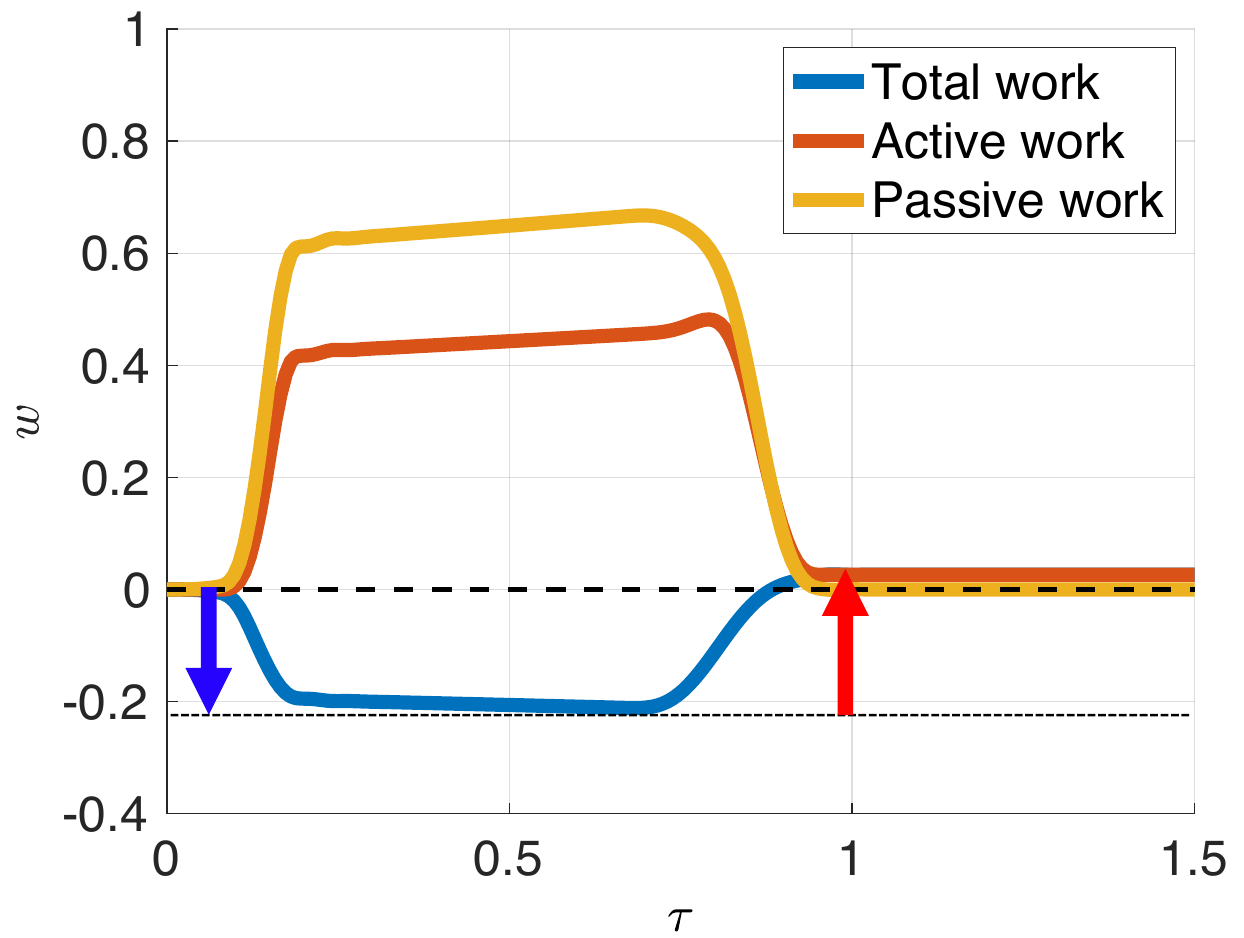}}
        \caption{Work curves by EGJ wall for $\gamma=0.06$}
        \label{fig:FastOpenWork}
    \end{subfigure}
    \hfill
    \begin{subfigure}[b]{0.48\textwidth}  
        \centering 
        {\includegraphics[trim=20 190 40 190,clip,width=\textwidth]{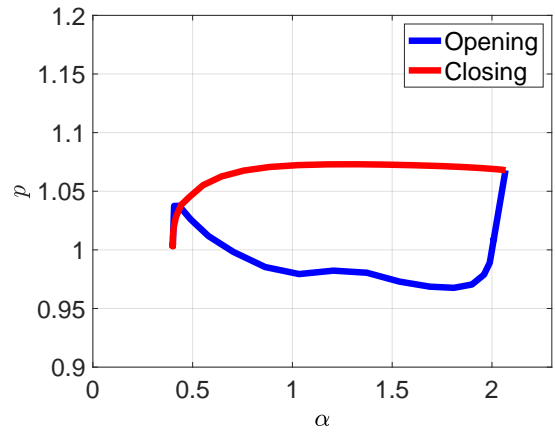}}
        \caption{Pressure-area loop at EGJ for $\gamma=0.06$}
        \label{fig:FastOpenLoop}
    \end{subfigure}
     \hfill
         \begin{subfigure}[b]{0.48\textwidth}  
        \centering 
        {\includegraphics[trim=20 130 40 140,clip,width=\textwidth]{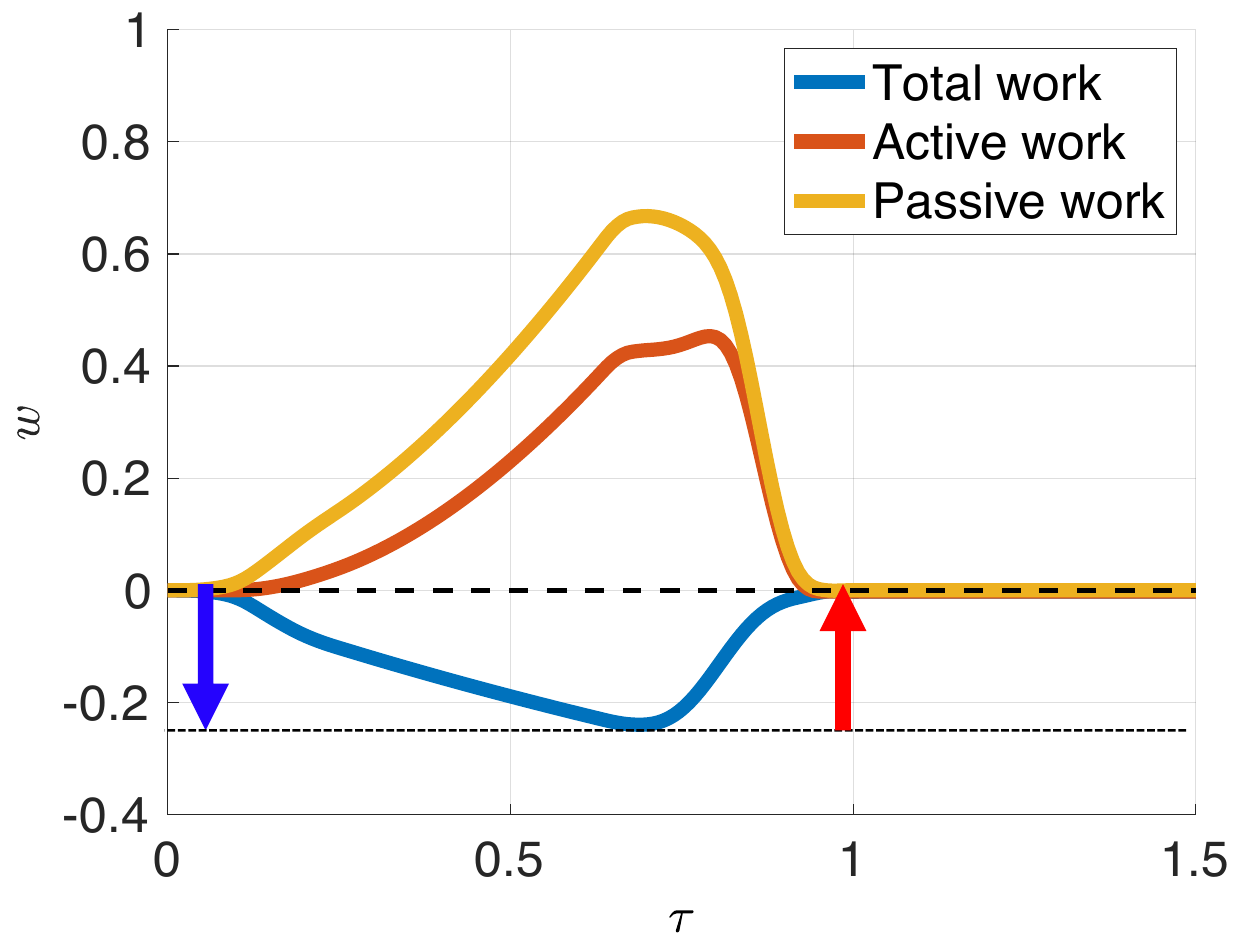}}
        \caption{Work curves by EGJ wall for $\gamma=0.6$}
        \label{fig:SlowOpenWork}
    \end{subfigure}
     \hfill
    \begin{subfigure}[b]{0.48\textwidth}  
        \centering 
        {\includegraphics[trim=20 190 40 190,clip,width=\textwidth]{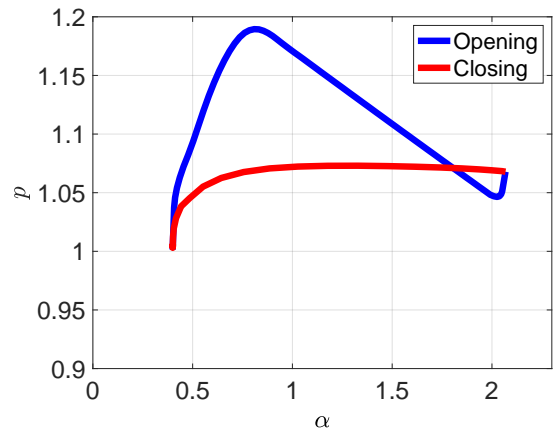}}
        \caption{Pressure-area loop at EGJ for $\gamma=0.6$}
        \label{fig:SlowOpenLoop}
    \end{subfigure}
    \caption{Comparing two simulations with different $\gamma$ values.} 
    \label{fig:OpeningSpeedComp}
\end{figure*}

While the conclusions obtained in sections \ref{Tube_Stiffness} and \ref{Fluid_Viscosity}  are theoretically meaningful, it is important to note that fluid viscosity and esophagus stiffness are mostly consistent among clinical data. Therefore, they cannot be considered the main reasons for the presence of the two loop types. The last parameter discussed in this writing is $\gamma$, which dictates the EGJ relaxation speed. Figure \ref{fig:OpeningSpeedComp} presents the work plots and pressure area loops of two different contractile cycle simulations with the same input parameters except for the value of $\gamma$. Figures \ref{fig:FastOpenWork} and \ref{fig:FastOpenLoop} display the work curves and pressure-area loop, respectively, for the simulation where the EGJ relaxes quickly. Figures \ref{fig:SlowOpenWork} and \ref{fig:SlowOpenLoop} display the work curves and pressure-area loop, respectively, for the simulation where the EGJ relaxes slowly. As these figures show, the EGJ relaxation speed alone can dictate the loop type, with tone dominant loop associated with fast relaxation of the EGJ and pressure dominant loop associated with slow relaxation of the EGJ.

In order to explain why the relaxation speed affects the loop type, we again look at the work curves. Starting with the total work curve, we see from figures \ref{fig:FastOpenWork} and \ref{fig:SlowOpenWork} that the magnitude of the total work done by the EGJ wall to close the EGJ ($w_{\text{close}}$) is equal in both cases (the red arrows in figures \ref{fig:FastOpenWork} and \ref{fig:SlowOpenWork} have the same length). This is because the closing pattern is the same in both cases. However, this is not case for $w_{\text{open}}$ (blue arrows in figures \ref{fig:FastOpenWork} and \ref{fig:SlowOpenWork} are of different length). The total work needed to open the EGJ when the EGJ relaxes slowly is greater than when the EGJ relaxes quickly. Therefore, the difference must originate from the opening work, where $|w_{\text{open,slow}}|>|w_{\text{open,fast}}|$. 

The total work done by the wall is the sum of its active and passive components. Hence, we will now examine each of these curves separately, starting with the passive work. The maximum magnitude of passive, elastic energy stored in the EGJ wall is equal because in both cases the wall expands to the same maximum cross-sectional area, and the work to achieve that is independent of path. To better understand this, think of an elastic band of length $L_o$ that is being stretched to some final length $L_f$, where $L_f>L_o$. It does not matter whether the band is being stretched quickly or slowly, as long as it is stretched to the same final length, it stores the same amount of passive energy.

Consequently, the difference in $w_{\text{open}}$ must depend on the active work done by the EGJ wall. As figures \ref{fig:FastOpenWork} and \ref{fig:SlowOpenWork} show, the maximum magnitude of the active work in the two examples is not the same, with the fast EGJ relaxation case having higher magnitude. Recall that when the EGJ starts relaxing, the reference area at the EGJ starts increasing such that the actual shape lags the relaxed shape. The active work is the work done by the EGJ tone to overcome fluid suction and match the actual shape to the relaxed shape. In parallel to relaxation, the peristaltic wave that enters the domain and starts traveling down the tube length increases the pressure at the EGJ location. The increase in pressure reduces the amount of active work the EGJ wall needs to exert in two ways. First, it reduces the effect of fluid suction. Second, it increases the cross-sectional area at the EGJ, making the actual EGJ cross-section closer to the reference cross-section. However when the EGJ relaxes quickly, the reference area grows quickly, but the peristaltic wave has barely traveled so there is less pressure developed in the tube. Hence, the EGJ tone must apply more active work to increase the actual EGJ cross-sectional area to match the fast-increasing reference area.

Note that towards the end of the contractile cycle, the EGJ wall contracts into its original shape, and therefore the EGJ wall contraction speed at the end of the contractile cycle (EGJ closes) is also a parameter. However, while we recognize that the EGJ contraction speed is a parameter in the system, relaxation pattern is the more dominant parameter in differentiating loop types. This is because contraction speed does not widely vary across subjects. By looking at clinical FLIP recordings of different contractile cycles, the active contraction of the EGJ is set by the peristaltic wave, similar to the pattern used in the model above \cite{AcharyaEsoWork2020, Lin2013,Carlson2016}. Therefore, the EGJ contraction speed depends on the speed of the traveling wave. The wave speed $c$ is a known constant ($c=1.5-3\ \text{cm/s}$) that was determined by calculating the distance traveled by contraction wave over time \cite{Kou2015ajpgi,Li1994}. Hence, we assume that the EGJ contraction speed does not differentiate among subjects and cannot be the main parameter controlling loop type in clinical peristalsis.

\subsection{Application to Clinical Data} \label{Application_Clinical_Data}

Since the fluid viscosity and esophagus stiffness are generally the same among subjects, the results presented in section \ref{Parameters_Loop_Type} indicate that the main parameter causing the difference in the pressure-area loops is neurological. We wish to examine this conclusion by directly calculating and plotting the work curves of clinical data alongside their pressure-area loops. For this calculation, the cross-sectional area values are direct FLIP readings. The activation function ($\theta(x,t)$), $P_o$, and the ratio $K_e/A_o$ are calculated as proposed by \cite{Halder_2021} using the tube law relation in equation (\ref{eqn:tube_law}). The fluid properties are density $\rho=1000\ \text{kg/m}^{3}$ and viscosity $\mu=0.001\ \text{Pa}\cdot\text{s}$ \cite{Kou2015ajpgi}.

\begin{figure*}
    \centering{{\includegraphics[trim=0 0 0 0,clip,width=0.9\textwidth]{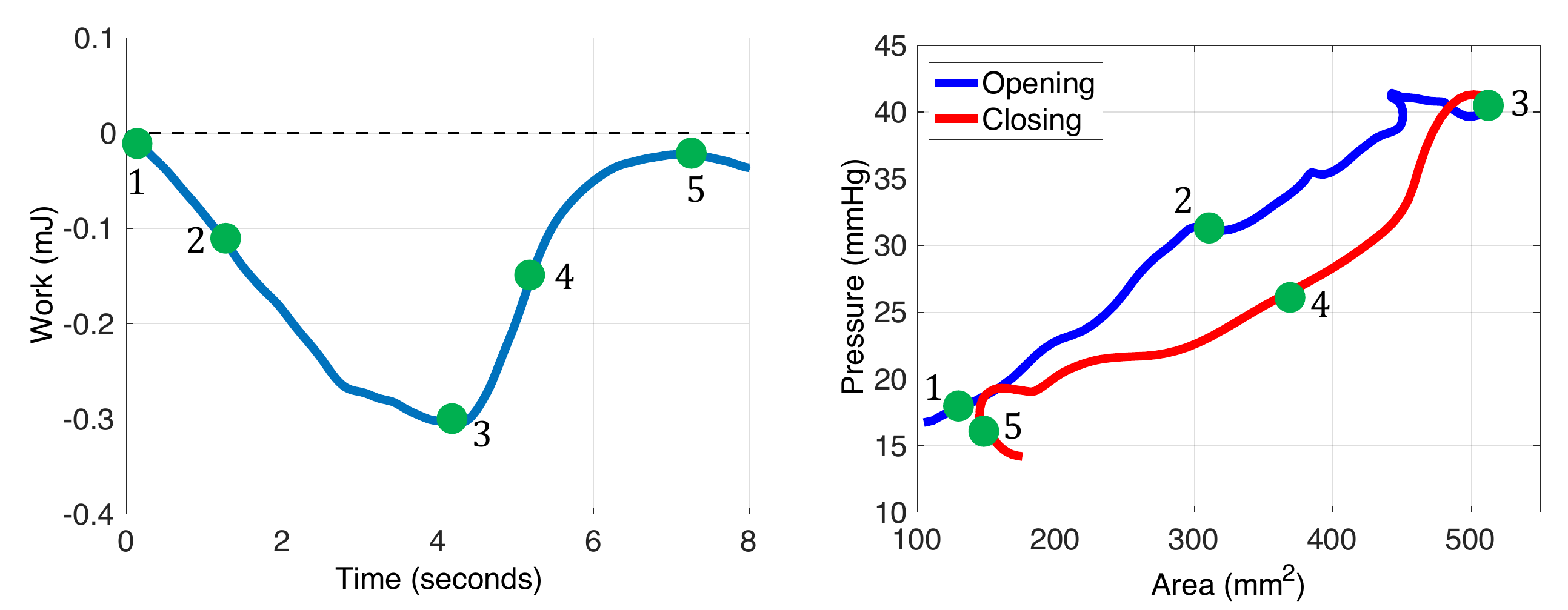}}}
    \caption{Tracking the work curve and the pressure-area loop at EGJ location of a clinical contractile cycle.}
    \label{fig:trackTimeC}
\end{figure*}

Figure \ref{fig:trackTimeC} presents the graph of the total work done by the EGJ wall (left) alongside the corresponding pressure-area loop at the EGJ location (right) of a clinical contractile cycle. Notice that the final work (at point 5) is negative, a characteristic of a pressure dominant loop, which is exactly the loop type corresponding to this work plot. Moreover, the overall pattern of the work curve and its relation to the loop is the same as the simulations, as observed by highlighting five time instances on the two figures. Hence, we can explain the clinical work curve based on the conclusions from the simulation results discussed in section \ref{Work_EGJ}.

\begin{figure*}
    \centering
    \begin{subfigure}[b]{0.9\textwidth}
        \centering
        {\includegraphics[trim=0 90 20 100,clip,width=\textwidth]{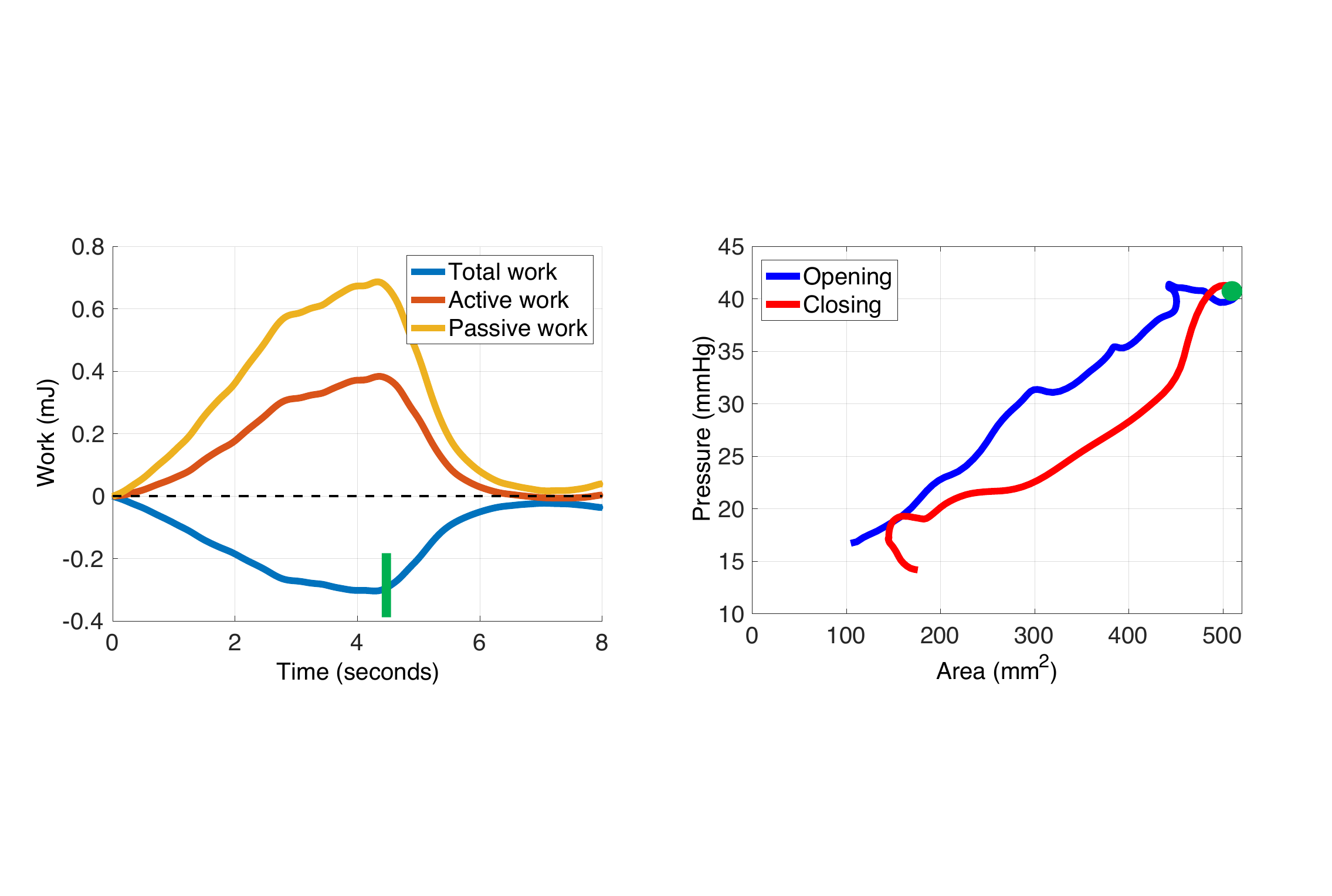}}
        \caption{Work curves and pressure-area loop extracted from clinical data. The graph on the left shows the total, passive, and active work done by the EGJ wall as a function of time. This work plot corresponds to a contractile cycle exhibiting a pressure dominant loop, as displayed on the right figure.}
        \label{fig:ClinicalPDLWork}
    \end{subfigure}
    \hfill
    \begin{subfigure}[b]{0.9\textwidth}  
        \centering 
        {\includegraphics[trim=0 90 20 100,clip,width=\textwidth]{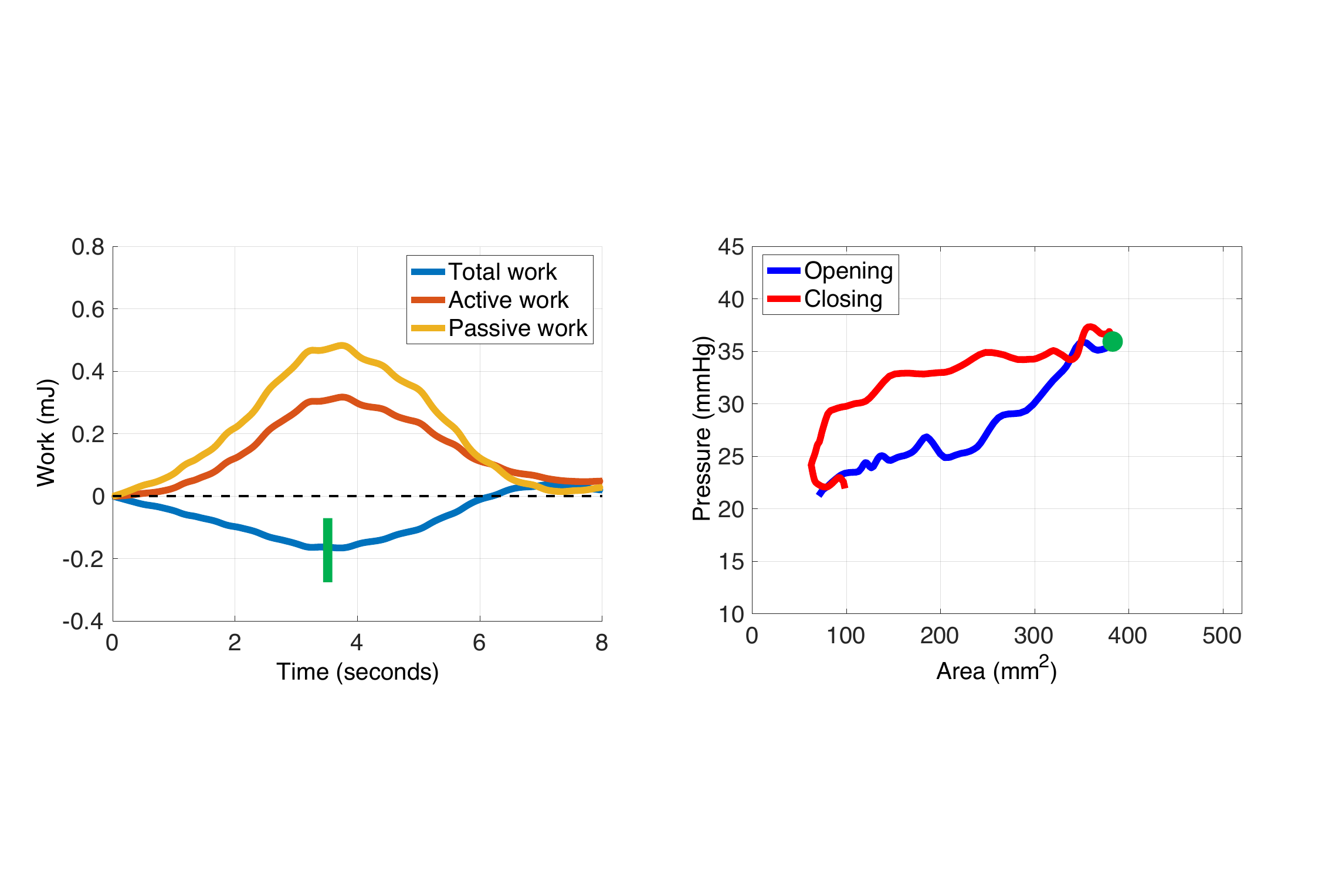}}
        \caption{Work curves and pressure-area loop extracted from clinical data. The graph on the left shows the total, passive, and active work done by the EGJ wall as a function of time. This work plot corresponds to a contractile cycle exhibiting a tone dominant loop, as displayed on the right figure.}
        \label{fig:ClinicalTDLWork}
    \end{subfigure}
       \caption{Comparing work plots and pressure-area loops of two clinical contractile cycles. The figure at the top corresponds to contractile cycle with pressure dominant loop whereas the figure at the bottom figure corresponds to contractile cycle with tone dominant loop} 
    \label{fig:clinicalComparison}
\end{figure*}

Next we wish to compare two clinical contractile cycles, one with a pressure dominant loop and the other with a tone dominant loop, by plotting their total, active, and passive work curves, as presented in figure \ref{fig:clinicalComparison}. As the figure shows, the final total work done by the EGJ wall throughout the entire cycle is positive for the tone dominant loop, and the passive, active, and total work curves look similar to the ones in the simulations. The most important observation however to extract from this figure is the green mark. This mark represents the time instance in which the EGJ is fully open. As the figure shows, the full opening for the tone dominant loop occurs earlier in the cycle than for the pressure dominant loop. As concluded in section \ref{Opening_Speed}, this is the parameter differentiating the two loops. The slower the opening, the more effect the pressure has on the opening, and the less active work the EGJ needs to exert on the fluid in order to open.

\section{Conclusion} \label{Conclusion}

In this work, we dissected a clinical phenomenon into its physical parameters. Hence, rather than examining the pressure-area loops at the esophagogastric junction as a function of time, we aimed to get a better understanding of the EGJ function through mechanics, and therefore, identify the leading parameter that dictates the loop shape. From the simulation and parametric study presented in this writing, we concluded that the dominant reason for the presence of two pressure-area loop types at the EGJ during a contractile cycle is neurally controlled. The activation function applied by the esophagus muscles dictates the loop type rather than fluid or material properties. Moreover, this work is another example of a fundamental study which helps us explain clinical observations. While the pressure-area loops at the EGJ are generally viewed as noisy, they are proven to be a result of a clear, repeating pattern that we can physically explain. 

The main goal of this study was to get a better understanding of the opening and closing mechanism of the EGJ during a contractile cycle. Post processing clinical data led to identifying two major pressure-area loop types. The tone dominant loop, identified as the loop where the closing curve is above the opening curve, is characterized by low esophagus stiffness, high fluid viscosity, and fast relaxation of the EGJ tone. In contractile cycles with tone dominant loop, the leading source of energy in the opening and closing of the EGJ is the EGJ tone rather than fluid pressure increase from the peristaltic contraction wave. The pressure dominant loop, identified as the loop where the opening curve is above the closing curve, is characterized by high esophagus stiffness, low fluid viscosity, and slow relaxation of the EGJ tone. In contractile cycles with pressure dominant loop, the leading source of energy in the opening and closing cycle is the increase in fluid pressure due to the peristaltic wave rather than the EGJ tone relaxing or contracting. These conclusions give us a better view into the physical held of the system, and a deeper understanding of the EGJ function. These insights bring us a step closer to identifying the underlying conditions that cause EGJ degradation.

\bibliographystyle{asmejour}   
\bibliography{pap_bib} 

\end{document}